RU98-7-B
%
%
\headline{\hfil \folio}
\hoffset=0.5truein
\hsize=5.5truein
\vsize=8truein
%
\catcode`@=11                           
\newskip\ttglue
\def\ninefonts{%
   \global\font\ninerm=cmr9%
   \global\font\ninei=cmmi9%
   \global\font\ninesy=cmsy9%
   \global\font\nineex=cmex10%
   \global\font\ninebf=cmbx9%
   \global\font\ninesl=cmsl9%
   \global\font\ninett=cmtt9%
   \global\font\nineit=cmti9%
   \skewchar\ninei='177%
   \skewchar\ninesy='60%
   \hyphenchar\ninett=-1%
   \moreninefonts
   \gdef\ninefonts{\relax}}%
\def\moreninefonts{\relax}                      


\def\elevenfonts{%
   \global\font\elevenrm=cmr10 scaled \magstephalf%
   \global\font\eleveni=cmmi10 scaled \magstephalf%
   \global\font\elevensy=cmsy10 scaled \magstephalf%
   \global\font\elevenex=cmex10%
   \global\font\elevenbf=cmbx10 scaled \magstephalf%
   \global\font\elevensl=cmsl10 scaled \magstephalf%
   \global\font\eleventt=cmtt10 scaled \magstephalf%
   \global\font\elevenit=cmti10 scaled \magstephalf%
   \global\font\elevenss=cmss10 scaled \magstephalf%
   \skewchar\eleveni='177%
   \skewchar\elevensy='60%
   \hyphenchar\eleventt=-1%
   \moreelevenfonts
   \gdef\elevenfonts{\relax}}%
\def\moreelevenfonts{\relax}

\def\twelvefonts{%
   \global\font\twelverm=cmr10 scaled \magstep1%
   \global\font\twelvei=cmmi10 scaled \magstep1%
   \global\font\twelvesy=cmsy10 scaled \magstep1%
   \global\font\twelveex=cmex10 scaled \magstep1%
   \global\font\twelvebf=cmbx10 scaled \magstep1%
   \global\font\twelvesl=cmsl10 scaled \magstep1%
   \global\font\twelvett=cmtt10 scaled \magstep1%
   \global\font\twelveit=cmti10 scaled \magstep1%
   \global\font\twelvess=cmss10 scaled \magstep1%
   \skewchar\twelvei='177%
   \skewchar\twelvesy='60%
   \hyphenchar\twelvett=-1%
   \moretwelvefonts
   \gdef\twelvefonts{\relax}}%
\def\moretwelvefonts{\relax}                    

\def\fourteenfonts{%
   \global\font\fourteenrm=cmr10 scaled \magstep2%
   \global\font\fourteeni=cmmi10 scaled \magstep2%
   \global\font\fourteensy=cmsy10 scaled \magstep2%
   \global\font\fourteenex=cmex10 scaled \magstep2%
   \global\font\fourteenbf=cmbx10 scaled \magstep2%
   \global\font\fourteensl=cmsl10 scaled \magstep2%
   \global\font\fourteenit=cmti10 scaled \magstep2%
   \global\font\fourteenss=cmss10 scaled \magstep2%
   \skewchar\fourteeni='177%
   \skewchar\fourteensy='60%
   \morefourteenfonts
   \gdef\fourteenfonts{\relax}}%
\def\morefourteenfonts{\relax}                  


\def\tenmibfonts{
   \global\font\tenmib=cmmib10%
   \global\font\tenbsy=cmbsy10%
   \skewchar\tenmib='177%
   \skewchar\tenbsy='60%
   \gdef\tenmibfonts{\relax}}

\def\elevenmibfonts{
   \global\font\elevenmib=cmmib10 scaled \magstephalf%
   \global\font\elevenbsy=cmbsy10 scaled \magstephalf%
   \skewchar\elevenmib='177%
   \skewchar\elevenbsy='60%
   \gdef\elevenmibfonts{\relax}}

\def\twelvemibfonts{
   \global\font\twelvemib=cmmib10 scaled \magstep1%
   \global\font\twelvebsy=cmbsy10 scaled \magstep1%
   \skewchar\twelvemib='177%
   \skewchar\twelvebsy='60%
   \gdef\twelvemibfonts{\relax}}

\def\fourteenmibfonts{
   \global\font\fourteenmib=cmmib10 scaled \magstep2%
   \global\font\fourteenbsy=cmbsy10 scaled \magstep2%
   \skewchar\fourteenmib='177%
   \skewchar\fourteenbsy='60%
   \gdef\fourteenmibfonts{\relax}}

\def\mib{
   \tenmibfonts%
   \textfont0=\tenbf\scriptfont0=\sevenbf%
   \scriptscriptfont0=\fivebf%
   \textfont1=\tenmib\scriptfont1=\seveni%
   \scriptscriptfont1=\fivei%
   \textfont2=\tenbsy\scriptfont2=\sevensy%
   \scriptscriptfont2=\fivesy}%

\def\ninepoint{\ninefonts               
   \def\rm{\fam0\ninerm}%
   \textfont0=\ninerm\scriptfont0=\sevenrm\scriptscriptfont0=\fiverm
   \textfont1=\ninei\scriptfont1=\seveni\scriptscriptfont1=\fivei
   \textfont2=\ninesy\scriptfont2=\sevensy\scriptscriptfont2=\fivesy
   \textfont3=\nineex\scriptfont3=\nineex\scriptscriptfont3=\nineex
   \textfont\itfam=\nineit\def\it{\fam\itfam\nineit}%
   \textfont\slfam=\ninesl\def\sl{\fam\slfam\ninesl}%
   \textfont\ttfam=\ninett\def\tt{\fam\ttfam\ninett}%
   \textfont\bffam=\ninebf
   \scriptfont\bffam=\sevenbf
   \scriptscriptfont\bffam=\fivebf\def\bf{\fam\bffam\ninebf}%
   \def\mib{\relax}%
   \tt\ttglue=.5emplus.25emminus.15em
   \normalbaselineskip=11pt
   \setbox\strutbox=\hbox{\vrule height 8pt depth 3pt width 0pt}%
   \normalbaselines\rm\singlespaced}%

\def\tenpoint{
   \def\rm{\fam0\tenrm}%
   \textfont0=\tenrm\scriptfont0=\sevenrm\scriptscriptfont0=\fiverm
   \textfont1=\teni\scriptfont1=\seveni\scriptscriptfont1=\fivei
   \textfont2=\tensy\scriptfont2=\sevensy\scriptscriptfont2=\fivesy
   \textfont3=\tenex\scriptfont3=\tenex\scriptscriptfont3=\tenex
   \textfont\itfam=\tenit\def\it{\fam\itfam\tenit}%
   \textfont\slfam=\tensl\def\sl{\fam\slfam\tensl}%
   \textfont\ttfam=\tentt\def\tt{\fam\ttfam\tentt}%
   \textfont\bffam=\tenbf
   \scriptfont\bffam=\sevenbf
   \scriptscriptfont\bffam=\fivebf\def\bf{\fam\bffam\tenbf}%
   \def\mib{%
      \tenmibfonts%
      \textfont0=\tenbf\scriptfont0=\sevenbf%
      \scriptscriptfont0=\fivebf%
      \textfont1=\tenmib\scriptfont1=\seveni%
      \scriptscriptfont1=\fivei%
      \textfont2=\tenbsy\scriptfont2=\sevensy%
      \scriptscriptfont2=\fivesy}%
   \tt\ttglue=.5emplus.25emminus.15em
   \normalbaselineskip=12pt
   \setbox\strutbox=\hbox{\vrule height 8.5pt depth 3.5pt width 0pt}%
   \normalbaselines\rm\singlespaced}%

\def\elevenpoint{\elevenfonts           
   \def\rm{\fam0\elevenrm}%
   \textfont0=\elevenrm\scriptfont0=\sevenrm\scriptscriptfont0=\fiverm
   \textfont1=\eleveni\scriptfont1=\seveni\scriptscriptfont1=\fivei
   \textfont2=\elevensy\scriptfont2=\sevensy\scriptscriptfont2=\fivesy
   \textfont3=\elevenex\scriptfont3=\elevenex\scriptscriptfont3=\elevenex
   \textfont\itfam=\elevenit\def\it{\fam\itfam\elevenit}%
   \textfont\slfam=\elevensl\def\sl{\fam\slfam\elevensl}%
   \textfont\ttfam=\eleventt\def\tt{\fam\ttfam\eleventt}%
   \textfont\bffam=\elevenbf
   \scriptfont\bffam=\sevenbf
   \scriptscriptfont\bffam=\fivebf\def\bf{\fam\bffam\elevenbf}%
   \def\mib{%
      \elevenmibfonts%
      \textfont0=\elevenbf\scriptfont0=\sevenbf%
      \scriptscriptfont0=\fivebf%
      \textfont1=\elevenmib\scriptfont1=\seveni%
      \scriptscriptfont1=\fivei%
      \textfont2=\elevenbsy\scriptfont2=\sevensy%
      \scriptscriptfont2=\fivesy}%
   \tt\ttglue=.5emplus.25emminus.15em
   \normalbaselineskip=13pt
   \setbox\strutbox=\hbox{\vrule height 9pt depth 4pt width 0pt}%
   \normalbaselines\rm\singlespaced}%

\def\twelvepoint{\twelvefonts\ninefonts 
   \def\rm{\fam0\twelverm}%
   \textfont0=\twelverm\scriptfont0=\ninerm\scriptscriptfont0=\sevenrm
   \textfont1=\twelvei\scriptfont1=\ninei\scriptscriptfont1=\seveni
   \textfont2=\twelvesy\scriptfont2=\ninesy\scriptscriptfont2=\sevensy
   \textfont3=\twelveex\scriptfont3=\twelveex\scriptscriptfont3=\twelveex
   \textfont\itfam=\twelveit\def\it{\fam\itfam\twelveit}%
   \textfont\slfam=\twelvesl\def\sl{\fam\slfam\twelvesl}%
   \textfont\ttfam=\twelvett\def\tt{\fam\ttfam\twelvett}%
   \textfont\bffam=\twelvebf
   \scriptfont\bffam=\ninebf
   \scriptscriptfont\bffam=\sevenbf\def\bf{\fam\bffam\twelvebf}%
   \def\mib{%
      \twelvemibfonts\tenmibfonts%
      \textfont0=\twelvebf\scriptfont0=\ninebf%
      \scriptscriptfont0=\sevenbf%
      \textfont1=\twelvemib\scriptfont1=\ninei%
      \scriptscriptfont1=\seveni%
      \textfont2=\twelvebsy\scriptfont2=\ninesy%
      \scriptscriptfont2=\sevensy}%
   \tt\ttglue=.5emplus.25emminus.15em
   \normalbaselineskip=14pt
   \setbox\strutbox=\hbox{\vrule height 10pt depth 4pt width 0pt}%
   \normalbaselines\rm\singlespaced}%

\def\fourteenpoint{\fourteenfonts\twelvefonts 
   \def\rm{\fam0\fourteenrm}%
   \textfont0=\fourteenrm\scriptfont0=\twelverm\scriptscriptfont0=\tenrm
   \textfont1=\fourteeni\scriptfont1=\twelvei\scriptscriptfont1=\teni
   \textfont2=\fourteensy\scriptfont2=\twelvesy\scriptscriptfont2=\tensy
   \textfont3=\fourteenex\scriptfont3=\fourteenex
      \scriptscriptfont3=\fourteenex
   \textfont\itfam=\fourteenit\def\it{\fam\itfam\fourteenit}%
   \textfont\slfam=\fourteensl\def\sl{\fam\slfam\fourteensl}%
   \textfont\bffam=\fourteenbf
   \scriptfont\bffam=\twelvebf
   \scriptscriptfont\bffam=\tenbf\def\bf{\fam\bffam\fourteenbf}%
   \def\mib{%
      \fourteenmibfonts\twelvemibfonts\tenmibfonts%
      \textfont0=\fourteenbf\scriptfont0=\twelvebf%
      \scriptscriptfont0=\tenbf%
      \textfont1=\fourteenmib\scriptfont1=\twelvemib%
      \scriptscriptfont1=\tenmib%
      \textfont2=\fourteenbsy\scriptfont2=\tenbsy%
      \scriptscriptfont2=\tenbsy}%
   \normalbaselineskip=17pt
   \setbox\strutbox=\hbox{\vrule height 12pt depth 5pt width 0pt}%
   \normalbaselines\rm\singlespaced}%
%
%

\def\singlespaced{
   \baselineskip=\normalbaselineskip}           


%
%
\twelvepoint
%
%
\def\begintitle{\begingroup%
\obeylines\fourteenpoint\bf\parindent=0.29truein}
\def\endtitle{\vglue 1truecm\endgroup}

\def\showheadline#1#2{\headline={\ifnum\pageno>1{\ifodd\pageno{\hfil\tenpoint #1\hfil} %
\else{\hfil\tenpoint #2\hfil}\fi} \else{\hfil}\fi}}

\def\beginauthor{\begingroup%
\obeylines\fourteenpoint\parindent=0.29truein}
\def\endauthor{\endgroup}

\def\address#1{\hbox to \hsize{\hglue 0.29in\relax
\vbox{\hsize=4.70in\relax\rightskip=0pt plus 1in\relax\noindent#1}\hfil}}

\long\def\beginaddress#1\endaddress{\vglue 6pt\address{#1}\vglue 24pt}

\def\beginabstract{\begingroup\leftskip=0.29in%
\tenpoint\noindent{\bf Abstract\ \ \ }}
\def\endabstract{\vskip 1pt minus1pt\endgroup}

\def\finalversion{\headline{\hfil}}

\def\section#1{\vskip 24pt plus4pt minus4pt\goodbreak\leftline{\bf #1}%
\vglue 12pt\nobreak\noindent\kern -0.0em}

\def\subsection#1{\vskip 12pt plus4pt minus4pt\goodbreak\leftline{\bf #1}%
\nobreak\noindent\kern -0.0em}

\def\subsubsection#1{\vskip 12pt plus4pt minus4pt\goodbreak\leftline{\it #1}%
\nobreak\noindent\kern -0.0em}

\def\tablerule{
\noalign{\vglue 6pt plus1pt minus1pt}
\noalign{\hrule height 0.4pt depth 0.3pt\relax}
\noalign{\vglue 6pt plus1pt minus1pt}
}

\def\begintable#1#2{\begingroup%
\vskip 12pt plus4pt minus 4pt\relax\goodbreak
\centerline{\tenpoint\noindent#1\ \ \ #2}
\vglue 12pt\nobreak
\hrule height 0.4pt depth 0.3pt\relax
\vglue 6pt\nobreak
}
\def\endtable{
\vglue 6pt\nobreak
\hrule height 0.4pt depth 0.3pt\relax
\vglue 12pt\nobreak
\vskip 12pt plus4pt minus 4pt\relax\goodbreak
\endgroup}

\def\begincaption#1{\begingroup\tenpoint\noindent#1\ \ \ }
\def\endcaption{\endgroup}

\newbox\@capbox                                 
\newcount\@caplines                             

\def\references{\section{REFERENCES}\tenpoint\parindent=0pt
\raggedright\rightskip=0pt plus 5em}

\def\ref#1#2{\hbox to \hsize{\vbox{\tenpoint\hsize=0.2in\relax #1\hfil}
\hfil\vtop{\hsize=4.75in\relax\tenpoint #2}}}
%
%
\vglue 1.0truein

\finalversion
\overfullrule 0pt
\input epsf
\begintitle
Path Integrals, BRST Identities
and Regularization Schemes
in Nonstandard Gauges
\endtitle

\beginauthor
Hai-cang Ren
\endauthor

\beginaddress
Department of Physics, The Rockefeller University,
New York, NY 10021, U. S. A. (permanent address)
\endaddress

\beginaddress
National Center of Theoretical Science,
National Tsing Hua University,
Hsinchu, Taiwan, ROC
\endaddress

\beginabstract
The path integral of a gauge theory is studied in Coulomb-like gauges. 
The Christ-Lee terms of operator ordering are reproduced {\it{within}} 
the path integration framework. In the presence of fermions, a new 
operator term, in addition to that of Christ-Lee, is discovered. Such 
kind of terms is found to be instrumental in restoring the invariance 
of the effective Lagrangian under a field dependent gauge transformation,
which underlies the BRST symmetry. A unitary regularization scheme which 
maintains manifest BRST symmetry and is free from energy divergences 
is proposed for a nonabelian gauge field. 

\endabstract

\section{I. Introduction}

The quantization of a gauge field in the continuum requires gauge fixing. 
The standard gauge for perturbative calculations is the covariant gauge, 
in which the theory can be regularized and renormalized systematically. 
On the other hand, noncovariant gauges, though computationally more involved,
possess a number of advantages of their own. 
Take Coulomb gauge as an example, all degrees of freedom there are 
physical and unitarity is manifest. The Coulomb propagator alone gives 
rise to the renormalization of the coupling constant [1] [2]. 
In addition, the explicit solubility of the Gauss law constraint in 
Coulomb gauge might make it easier to construct variational wave functional 
to explore the nonperturbative physics of a nonabelian gauge field.

The difficulties with noncovariant gauges of a nonabelian gauge theory
include 1) the complication of the curvilinear coordinates; 2) the lack of 
manifest BRST and Lorentz invariances; 3) the additional divergence, the 
energy divergence, because of the instantaneous nature of the bare Coulomb 
propagator and the ghost propagator; 
4) the lack of a unitary regularization scheme which can be applied to 
all orders and therefore the uncertainty of the renormalizability.

The operator ordering associated with the curvilinear coordinates 
has been investigated by Christ and Lee [3]. They 
derived the path integral in a noncovariant 
gauge from the Weyl ordering of the corresponding Hamiltonian 
and found that the effective Lagrangian contains
two nonlocal terms, referred to as Christ-Lee terms, in addition to the 
classical Lagrangian and the ghost determinant. These terms start to show 
up at the two loop order and their diagrammatic implications have been 
discussed to that order [4][5]. 
Here, I approach the noncovariant gauge strictly {\it{within}} the 
path integral formulation, along the line of Refs. 6 and 7. Starting with 
the discrete time path integral in the time 
axial gauge, the change of the integration variables to other gauges 
and the continuous time 
limit are examined carefully and the Christ-Lee terms are 
reproduced. In the presence of quark fields, a new nonlocal term 
of the effective Lagrangian involving fermion bilinears, which 
starts to show up at the {\it{one loop}} order is discovered. 

According to Feynman [8], a path integral of a quantum mechanical system 
is the $\epsilon\to 0$ limit of a multiple integral over canonical coordinates 
on a one dimensional lattice of time slices separated by $\epsilon$. 
The exponent of the weighting factor on each time slice is equal to 
$i\epsilon$ times the classical Lagrangian only if the canonical coordinates 
are cartesian [2]. This is the case of the time-axial gauge of a gauge theory.
Even there, the velocities in the Lagrangian is the {\it mean} velocity between 
the neighboring time slices instead of instantaneous ones. When transforming 
the integration variables to curvilinear ones, or to other gauges, e.g. 
Coulomb gauge, one has to keep track of all the contributions to the limit 
$\epsilon\to 0$, which introduces the Christ-Lee operator ordering terms in 
addition to the classical Lagrangian in terms of the new coordinates and the 
corresponding jacobian. For the same reason, 
the classical Lagrangian with the mean velocities is not exactly invariant 
under a gauge transformation which depends on canonical coordinates and 
the variation contributes to the limit $\epsilon\to 0$. 
The standard form of the BRST identity is only recovered after 
including the Christ-Lee terms. Alternatively, one may retain a nonvanishing 
$\epsilon$ and this lends us to a gauge theory with discrete time 
coordinates, which is manifest BRST invariant.
For a field theory, this discrete time formulation serves as a unitary 
regularization scheme, which regularizes the energy divergence and the 
ordinary ultraviolet divergence at one shot. It also possesses several 
technical advantages which may be helpful in higher orders.

This paper is organized as follows: I will illustrate the technique of 
the gauge fixing within the path integral formulation and the derivation 
of the BRST identity of the soluble model of Ref. [5] in the next two 
sections. The application to a nonabelian gauge field is discussed in 
the Sections IV and V. There I will also test the discrete time regularization 
scheme by evaluating the one loop correction to the Coulomb propagator. 
The comparison of my regularization scheme with others and 
some comments on the renormalizability will be discussed in the 
final section. Except for the fermionic 
operator ordering term, the interplay between the BRST identity and 
the operator ordering terms and the discrete time regularization scheme, 
all other results are not new. But it is instructive to see how the 
operator ordering terms come about without referring to the operator 
formulations. It is also amazing to see how similar in formulation the 
soluble model and the nonabelian gauge field are.

\section{II. The Path Integral of a Soluble Model}

	The soluble model proposed by Friedberg, Lee, Pang and the author 
[5] provides a playground for the investigation of gauge fixing 
and the BRST invariance in some nonstandard gauges within the 
path integral formulation. 

	The Lagrangian of the soluble model is 
$$L={1\over 2}[(\dot x+g\xi y)^2+(\dot y-g\xi x)^2+(\dot z-\xi)^2]
-U(x^2+y^2).\eqno(2.1)$$ It is invariant under the following 
gauge transformation
$$x\to x^\prime=x\cos\alpha-y\sin\alpha,\eqno(2.2)$$
$$y\to y^\prime=x\sin\alpha+y\cos\alpha,\eqno(2.3)$$
$$z\to z^\prime=z+{1\over g}\alpha\eqno(2.4)$$
and $$\xi\to \xi^\prime=\xi+{1\over g}\dot\alpha\eqno(2.5)$$
with $\alpha$ an arbitrary function of time. The Lagrangian (2.1)
does not contain the time derivative of $\xi$ and the corresponding 
equation of motion reads
$${\partial L\over \partial\xi}=g[y(\dot x+g\xi y)-x(\dot y-g\xi x)]
-\dot z+\xi=0,\eqno(2.6)$$ which is the analog of the Gauss law of
a gauge field. In the following, 
we shall review the canonical quantization in the time-axial gauge, 
i. e., $\xi=0$, convert it into a path integral and transform carefully 
the path integral into the $\lambda$-gauge, i. e., $z=\lambda x$, 
an analog of the Coulomb gauge. 

In the time-axial gauge, the Lagrangian (2.1) becomes
$$L={1\over 2}(\dot X^2+\dot Y^2+\dot Z^2)-U(X^2+Y^2).\eqno(2.7)$$
The canonical momenta corresponding to $X$, $Y$, and $Z$ are 
$$P_X=\dot X=-i{\partial\over\partial X},\eqno(2.8)$$
$$P_Y=\dot Y=-i{\partial\over\partial Y}\eqno(2.9)$$ and 
$$P_Z=\dot Z=-i{\partial\over\partial Z},\eqno(2.10)$$
and the Hamiltonian operator reads
$$H={1\over 2}(P_X^2+P_Y^2+P_Z^2)+U(X^2+Y^2).\eqno(2.11)$$
The physical states in the Hilbert space are subject to the 
Gauss law constraint, i.e.
$$[P_Z+g(XP_Y-YP_X)]\vert>=0,\eqno(2.12)$$ as follows from (2.6) and 
the operator $P_Z+g(XP_y-YP_X)$ commutes with $H$. 
In terms of polar coordinates, $X=\rho\cos\Phi$ and $Y=\rho\sin\Phi$,
the wave function of a physical state takes the form
$$<X,Y,Z\vert>=\Psi(\rho, \Phi-gZ).\eqno(2.13)$$ For a harmonic oscillator
potential, $U={1\over 2}\omega^2(X^2+Y^2)$, the energy spectrum is given by
$$E=\omega(n_++n_-+1)+{1\over 2}g^2(n_+-n_-)^2\eqno(2.14)$$ with 
$n_+$, $n_-$ non-negative integers, and the corresponding eigenfunction 
can be expressed in terms of Laguerre polynomials. 

Following Feynman, the transition matrix element 
$<X,Y,Z\vert e^{-iHt}\vert >$ can be cast
into a path integral $$<X,Y,Z\vert e^{-iHt}\vert >=\lim_{\epsilon\to 0}
\Big({1\over 2i\pi\epsilon}\Big)^{{3\over 2}N}\int\prod_{n=0}^{N-1}
dX_ndY_ndZ_n$$ $$\times e^{i\epsilon\sum_{n=0}^{N-1}L(n)}<X_0,Y_0,Z_0\vert>,
\eqno(2.15)$$ where $\epsilon=t/N$ and 
$$L(n)={1\over 2}(\dot X_n^2+\dot Y_n^2+\dot Z_n^2)-U(X_n^2+Y_n^2)
\eqno(2.16)$$ with $\dot X_n=(X_{n+1}-X_n)/\epsilon$ etc.. In the rest 
part of the paper, the limit sign and the normalization factors like 
$(2i\pi\epsilon)^{-{3\over 2}N}$ will not be displayed explicitly.

As was pointed out in [3] and [5], the path integral (2.15) picks up the 
velocity $$(\dot X_n, \dot Y_n, \dot Z_n)\hbox{ as large as } 
O\Big(\epsilon^{-{1\over 2}}\Big).\eqno(2.17)$$
In other words, the contribution to the path 
integral comes from paths which can be more zigzag than classical 
ones. This has to be taken into account 
in variable transformations. The magnitudes of $X_n$, $Y_n$ and $Z_n$,
on the other hand, remains of the order one with a well defined 
initial wave function $<X_0,Y_0,Z_0\vert>$. To transform the path 
integral (2.15) to $\lambda$-gauge, i.e., $z=\lambda x$, we insert the 
identity $$1={\rm{const.}}\int\prod_{n=0}^{N-1}d\theta_n{\cal J}_n\delta
(z_n-\lambda x_n),\eqno(2.18)$$ with 
$$x_n=X_n\cos\theta_n-Y_n\sin\theta_n,\eqno(2.19)$$
$$y_n=X_n\sin\theta_n+Y_n\cos\theta_n,\eqno(2.20)$$
$$z_n=Z_n+{1\over g}\theta_n\eqno(2.21)$$ and $${\cal J}_n=
{1\over g}+\lambda y_n,\eqno(2.22)$$ we have  
$$<X,Y,Z\vert e^{-iHt}\vert >={\rm{const.}}\int\prod_{n=0}^{N-1}
dX_ndY_ndZ_nd\theta_n{\cal J}_n\delta(z_n-\lambda x_n)\times$$ $$\times 
e^{i\epsilon\sum_{n=0}^{N-1}L(n)}<X_0,Y_0,Z_0\vert>,\eqno(2.23)$$
Introducing back $\xi_n$ via $$\xi_n={1\over g}\dot\theta_n
={\theta_{n+1}-\theta_n\over g\epsilon}\eqno(2.24)$$ and changing the 
integration variables from $X_n$, $Y_n$, $Z_n$ and $\theta_n$ to $x_n$, 
$y_n$, $z_n$ and $\xi_n$, we obtain that
$$<X,Y,Z\vert e^{-iHt}\vert >={\rm{const.}}\int\prod_{n=0}^{N-1}
dx_ndy_ndz_nd\xi_n\delta(z_n-\lambda x_n)\times$$ $$\times 
e^{i\epsilon\sum_{n=0}^{N-1}L^\prime(n)}<x_0,y_0,z_0\vert>,\eqno(2.25)$$ 
where $$L^\prime(n)=L(n)-{i\over\epsilon}\ln{\cal J}_n\eqno(2.26)$$
with $L(n)$ the same Lagrangian (2.16). Written in terms of the new 
variables, $L(n)$ becomes
$$L(n)={1\over 2\epsilon^2}(\tilde r_{n+1}e^{-i\epsilon g\xi_n\sigma_2}
-\tilde r_n)(e^{i\epsilon g\xi_n\sigma_2}r_{n+1}-r_n)
+{1\over 2}(\dot z_n-\xi_n)^2-U(\tilde r_n r_n),\eqno(2.27)$$
where we have grouped $x_n$ and $y_n$ into a $2\times 1$ matrix 
$$r_n=\left(\matrix{x_n\cr y_n}\right)\eqno(2.28)$$
and $\sigma_2$ is the second Pauli matrix.

For finite $\epsilon$, (2.25) with (2.26)-(2.27) define a one-dimensional 
lattice gauge field and the limit $\epsilon\to 0$ corresponds to its 
continuum limit. As $\epsilon\to 0$, it follows from (2.17), 
(2.19)-(2.21) and (2.24) that 
$$(\dot x_n, \dot y_n, \dot z_n, \xi_n)=O(\epsilon^{-{1\over 2}}).
\eqno(2.29)$$ Several terms beyond the naive continuum limit 
have to be kept when expanding the exponential $e^{i\epsilon g\xi_n\sigma_2}$
in (2.27) according to $\xi_n$ [3]. In addition, the commutativity between time derivative and path 
integral contractions requires the Lagrangian to be written in terms of 
$\dot x_n$, $\dot y_n$, $\dot z_n$, $\xi_n$, $\bar x_n$, $\bar y_n$ and 
$\bar z_n$ with $\bar x_n=(x_n+x_{n+1})/2$ etc.. In another word, we write
$$L^\prime(n)\equiv L^\prime(\dot x_n,\dot y_n,\dot z_n,\xi_n,x_n, 
y_n,z_n)=L^\prime(\dot x_n,\dot y_n,\dot z_n,\xi_n,\bar x_n,  
\bar y_n,\bar z_n)+\delta L^\prime(n)\eqno(2.30)$$ with 
$$\delta L^\prime(n)=L^\prime(\dot x_n,\dot y_n,\dot z_n,\xi_n,x_n,
y_n,z_n)-L^\prime(\dot x_n,\dot y_n,\dot z_n,\xi_n,\bar x_n,\bar y_n,
\bar z_n).\eqno(2.31)$$ According to the estimate (2.29), the 
contribution from the potential energy $U$ to the difference (2.31) 
vanishes in the limit $\epsilon\to 0$. But that from the kinetic 
energy and from the jacobian (2.22) {\it{do not}}. With this precaution,
we rewrite $L^\prime(n)$ as
$$L^\prime(n)=L_{\rm{eff.}}(n)+{i\over 2\epsilon}(\ln{\cal J}_{n+1}
-\ln{\cal J}_n),\eqno(2.32)$$ where
$$L_{\rm{eff.}}(n)=L(n)-{i\over\epsilon}\ln\bar{\cal J}_n+{i\over\epsilon}
\Big[\ln\bar{\cal J}_n-{1\over 2}
(\ln{\cal J}_{n+1}+\ln{\cal J}_n)\Big]\eqno(2.33)$$ with
$$\bar{\cal J}_n={1\over g}+\lambda\bar y_n.\eqno(2.34)$$
The path integral (2.25) becomes then
$$<X,Y,Z\vert e^{-iHt}\vert >={\rm{const.}}{\cal J}_N^{-{1\over 2}}
\int\prod_{n=0}^{N-1}dx_ndy_ndz_nd\xi_n\delta(z_n-\lambda x_n)\times$$ 
$$\times e^{i\epsilon\sum_{n=0}^{N-1}L_{\rm{eff.}}(n)}
{\cal J}_0^{1\over 2}<x_0,y_0,z_0\vert>,\eqno(2.35)$$ 
Now it is the time to take the limit $\epsilon\to 0$. 
The small $\epsilon$ expansion of $L_{\rm{eff.}}(n)$ reads
$$L_{\rm{eff.}}(n)=L_{\rm{cl.}}(n)-{i\over\epsilon}\ln
\bar{\cal J}_n+\Delta L(n)+O(\epsilon^{1\over 2}),\eqno(2.36)$$ where
$$L_{\rm{cl.}}(n)={1\over 2}[(\dot x_n+g\xi_n\bar x_n)^2+
(\dot y_n-g\xi_n\bar y_n)^2+(\dot z_n-\xi_n)^2]-U(\bar x_n^2+\bar y_n^2)
\eqno(2.37)$$ is the classical Lagrangian but with {\it mean}
velocities and $$\Delta L(n)=-{1\over 8}\epsilon^2g^2\xi_n^2(
\tilde{\dot r_n}-ig\xi_n\tilde{\bar r_n}\sigma_2)\Big(\dot r_n
+{i\over 3}g\sigma_2\xi_n\bar r_n\Big)+{i\over 8}\epsilon
{\lambda^2g^2\dot y_n^2\over (1+\lambda g\bar y_n)^2}.\eqno(2.38)$$
The first term of $\Delta L(n)$ comes from the kinetic energy, the
second term from the jacobian. It follows from (2.29) that $\Delta L(n)
=O(1)$ and the terms not displayed all vanish as $\epsilon\to 0$.

It remains to convert $\Delta L(n)$ into an equivalent potential 
(to eliminate the explicit $\epsilon$ dependence). The recipe was given 
by Gervais and Jevicki [6], which we shall outline here. 
We assume that the integrations on the time slices $t=0$, $\epsilon$, 
$2\epsilon$, ..., $(n-1)\epsilon$ have been carried out and we are left 
with $$<X,Y,Z\vert e^{-iHt}\vert >={\rm{const.}}{\cal J}_N^{-{1\over 2}}
\int\prod_{m=n+1}^{N-1}dx_mdy_mdz_md\xi_m$$ $$\times\delta(z_m-\lambda x_m) 
e^{i\epsilon\sum_{m=n+1}^{N-1}L_{\rm{eff.}}(m)}$$
$$\times\int dx_ndy_ndz_nd\xi_n{\cal J}_n^{1\over 2}\delta(z_n-\lambda x_n)
e^{i\epsilon L_{\rm{cl.}}(n)}[1+i\epsilon\Delta L(n)]<x_n,y_n,z_n\vert e^
{-in\epsilon H}\vert>,\eqno(2.39)$$ where the corresponding transition matrix 
element, $<x_n,y_n,z_n|e^{-in\epsilon H}|>$ is a smooth function 
of $x_n$, $y_n$ and $z_n$. The structure of $\Delta L(n)$ is 
$$\Delta L(n)=\sum_lC_l(\bar x_n,\bar y_n)P_l(n)\epsilon^{{n_l\over 2}}
\eqno(2.40)$$ with $P_l(n)$ a product of $\dot x_n$, $\dot y_n$ and $\xi_n$ 
and $n_l$ the number of factors. Changing the 
integration variables from $x_n$, $y_n$ and $\xi_n$ to $\dot x_n$,
$\dot y_n$ and $\xi_n$ while replacing ($x_n$, $y_n$) by 
($x_{n+1}-\epsilon \dot x_n$, $y_{n+1}-\epsilon \dot y_n$) and 
($\bar x_n$, $\bar y_n$) by ($x_{n+1}-\epsilon \dot x_n/2$, 
$y_{n+1}-\epsilon \dot y_n/2$), we have,  
upon a Taylor expansion in terms of $\epsilon \dot x_n$ and 
$\epsilon\dot y_n$, that 
$$(2.39)={\rm{const.}}{\cal J}_N^{-{1\over 2}}\int\prod_{m=n+1}^{N-1}
dx_mdy_mdz_nd\xi_m\delta(z_m-\lambda x_m) 
e^{i\epsilon\sum_{m=n+1}^{N-1}L_{\rm{eff.}}(m)}$$
$$\int dx_ndy_ndz_nd\xi_n{\cal J}_n^{1\over 2}\delta(z_n-\lambda x_n)
e^{i\epsilon L_{\rm{cl.}}(n)}[1-i\epsilon{\cal V}(\bar x_n,\bar y_n)
+O(\epsilon^{3\over 2})]$$ $$\times<x_n,y_n,z_n\vert e^{-in\epsilon H}\vert>,
\eqno(2.41)$$ where $${\cal V}(\bar x_n,\bar y_n)=
-<\Delta L(n)>_{\rm{Gauss}}=-\sum_lC_l(\bar x_n,\bar y_n)<P_l(n)>_
{\rm{Gauss.}}\eqno(2.42)$$ and the Gauss average $<...>_{\rm{Gauss}}$ is
defined to be $$<F(n)>_{\rm{Gauss}}={\int d\dot x_nd\dot y_nd\xi_ne^{i\epsilon
L_{\rm{cl.}}(n)}F(n)\over \int d\dot x_nd\dot y_nd\xi_ne^{i\epsilon
L_{\rm{cl.}}(n)}}\eqno(2.43)$$ while regarding $\bar x_n$ and $\bar y_n$
constants. Such a procedure is valid even if a linear term of 
$\dot x_n$, $\dot y_n$ and $\xi_n$ with coefficients of the order one
is added to $L_{\rm{cl.}}(n)$, as will be the case when external sources are
introduced to generate various Green's functions. Introduce a $3\times 1$ 
matrix $$\left(\matrix{\zeta_{1n} \cr\zeta_{2n} \cr\zeta_{3n}}\right)
=\left(\matrix{\dot x_n+g\bar y_n\xi_n \cr\dot y_n-g\bar x_n\xi_n
\cr\xi_n-\lambda\dot x_n}\right),\eqno(2.44)$$ we have
$$\left(\matrix{\dot x_n \cr\dot y_n \cr\xi_n}\right)=
{1\over 1+\lambda g\bar y_n}\left(\matrix{1&0&-g\bar y_n \cr\lambda
g\bar x_n&1+\lambda g\bar y_n&g\bar x_n \cr\lambda&0&1 \cr}\right)
\left(\matrix{\zeta_{1n} \cr\zeta_{2n} \cr\zeta_{3n}}\right).
\eqno(2.45)$$ It follows from (2.43) that
$$<\zeta_{in}\zeta_{jm}>_{\rm{Gauss}}={i\over\epsilon}\delta_{nm}\delta_{ij}
.\eqno(2.46)$$ Working out the Wick contractions in (2.42) according to 
(2.45) and (2.46), we end up with 
$${\cal V}(x,y)=-{g^2(2+3\lambda^2)+\lambda^3g^3y\over 
8(1+\lambda gy)^3}+{\lambda^2g^4x^2(1+\lambda^2)\over 8(1+\lambda gy)^4}
,\eqno(2.47)$$ which agrees with the result obtained via Weyl ordering.
The effective Lagrangian $L_{\rm{eff.}}(n)$ in (2.35) is then replaced by 
that of Christ-Lee type, i.e. $${\cal L}(n)=L_{\rm{cl.}}(n)-{i\over\epsilon}
\ln\bar{\cal J}_n-{\cal V}(n).\eqno(2.48)$$

Before closing this section, I would like to remark that the subtleties 
of the path integral depends strongly on the way in which the gauge 
condition is introduced. Consider a general linear gauge fixing with the 
insertion (2.18) replaced by 
$$1={\rm{const.}}\int\prod_{n=0}^{N-1}d\theta_n{\cal J}
e^{-i\epsilon{1\over 2a}(z_n-\lambda x_n-\kappa\dot\xi_n
)^2}\eqno(2.49)$$ with $a$ and $\kappa>0$ gauge parameters like $\lambda$.
This is the discrete version of the gauge fixing used in [9] and the 
$\lambda$-gauge, (2.18), corresponds to $a=0$ and $\kappa=0$. 
The analysis in Appendix A gives rise to the estimates in the table I for 
the typical magnitude of $\xi$ in the path integral with different choices 
of the gauge parameters.

\topinsert
\begintable{Table I.}{The typical magnitude of $\xi$}
\halign to \hsize{\tabskip 10pt plus 2in\relax
\hglue 10pt#\hfil&\hfil#\hfil&\hfil#\hfil\cr
	       & $\kappa=0$                   & $\kappa\neq 0$\cr
\tablerule
$a=0$          & $O(\epsilon^{-{1\over2}})$   & $O(1)$\cr
$a\neq 0$      & $O(\epsilon^{-{3\over2}})$   & $O(1)$\cr
}
\endtable
\endinsert

For $\kappa\neq 0$, the limit $\epsilon\to 0$ is trivial and the same 
estimates apply to the gauge fixing with $x_n$ and $z_n$ in (2.49) replaced 
by $\bar x_n$ and $\bar z_n$, an analog of the covariant gauge in a 
relativistic field theory. But the limit $\kappa\to 0$ with a continuous
time will entail higher degrees of energy divergence for individual 
diagrams.

\section{III. The BRST Identity of the Soluble Model}

There are two approaches to the BRST identity of the soluble 
model (2.1). One can prove BRST invariance by introducing ghost 
variables and establish the identity with external sources. 
One may also start with the Slavnov-Taylor identity [10] and construct the 
BRST identity afterwards. It turns out the former is 
more straightforward for the path integral (2.25) with the lattice 
Lagrangian (2.26) and (2.27), while 
the latter is more convenient with the Christ-Lee type of path
integral. We shall illustrate both approaches in the following.

\subsection{III.1 Prior to the $\epsilon$-expansion}

Introducing the ghost variables $c_n$ and $\bar c_n$ and an auxiliary 
field $b_n$, the path integral (2.25) can be cast into
$$<X,Y,Z\vert e^{-iHT}\vert>={\rm{const.}}\int\prod_{n=0}^{N-1}
dx_ndy_ndz_nd\xi_ndb_ndc_nd\bar c_n\times$$ $$\times 
e^{i\epsilon\sum_{n=0}^{N-1}L_{\rm{BRST}}(n)}<x_0,y_0,z_0\vert>
\eqno(3.1)$$ where
$$L_{\rm{BRST}}(n)={1\over 2\epsilon^2}(\tilde r_{n+1}
e^{-i\epsilon g\xi_n\sigma_2}-\tilde r_n)(e^{i\epsilon g\xi_n\sigma_2}
r_{n+1}-r_n)+{1\over 2}(\dot z_n-\xi_n)^2-U(\tilde r_nr_n)$$
$$+b_n(z_n-\lambda x_n)-\bar c_n(1+\lambda gy_n)c_n.\eqno(3.2)$$ 
The integration measure and the Lagrangian (3.2) is invariant under 
the following transformation
$$\delta r_n=-ig\theta_n\sigma_2 r_n,\eqno(3.3)$$
$$\delta z_n=\theta_n,\eqno(3.4)$$
$$\delta\xi_n=\dot\theta_n,\eqno(3.5)$$
$$\delta b_n=0,\eqno(3.6)$$
$$\delta c_n=0\eqno(3.7)$$ and $$\delta\bar c_n=s_nb_n\eqno(3.8)$$
with $\theta_n=s_nc_n$ and $s_n$ a Grassmann number.  
For $n$-independent $s_n\equiv s$, an operator $Q$ such that $\delta=sQ$ 
can be extracted. It is straightforward to show that $Q^2=0$ and the 
transformation (3.3)-(3.8) is of BRST type.

To establish the BRST identity, we introduce the generating functional 
of connected Green's functions,
$$e^{iW(J,\zeta,u,\eta,\bar\eta)}=\lim_{T\to\infty}
<\vert e^{-iHT}\vert>={\rm{const.}}\int\prod_{n=0}^N
dx_ndy_ndz_nd\xi_ndb_ndc_nd\bar c_n\times$$ $$\times <\vert x_N,y_N,z_N>
e^{i\epsilon\sum_n[L_{\rm{BRST}}(n)+L_{\rm{ext.}}(n)]}
<x_0,y_0,z_0\vert>,\eqno(3.9)$$ where $L_{\rm{ext.}}(n)$ stands for the 
source term, i.e.
$$L_{\rm{ext.}}(n)=\tilde J_nr_n+\zeta_nz_n+u_n\xi_n+\bar
\eta_nc_n+\bar c_n\eta_n,\eqno(3.10)$$ $|>$ denotes the ground state of the 
system, and the limits $\epsilon\to 0$
and $N\epsilon=T\to\infty$ are understood for the right hand side.
It follows from the transformations (3.3)-(3.8) that
$$<\delta L_{\rm{ext.}}(n)>_\eta=-ig\tilde J_n\sigma_2<c_nr_n>_\eta
+(\zeta_n-\dot u_n)<c_n>_\eta+<b_n>_\eta\eta_n=0.\eqno(3.11)$$
This is the prototype of the BRST identity and can be 
converted into various useful forms.

\subsection{III.2 After the $\epsilon$-expansion}

After integrating out the ghost variables and carrying out the 
$\epsilon$-expansion, the path integral (3.9) becomes 
$$e^{iW(J,\zeta,u,\eta,\bar\eta)}=\lim_{T\to\infty}
<\vert e^{-iHT}\vert>={\rm{const.}}\int\prod_{n=0}^N
dx_ndy_ndz_nd\xi_ndb_n\times$$ $$\times <\vert x_N,y_N,z_N>
e^{i\epsilon\sum_n[{\cal L}(n)+L_{\rm{ext.}}(n)]}
<x_0,y_0,z_0\vert>,\eqno(3.12)$$ where
$${\cal L}(n)=L_{\rm{cl.}}(n)+b_n(z_n-\lambda x_n)
-{i\over\epsilon}\ln\bar{\cal J}_n-{\cal V}(n)\eqno(3.13)$$ is the 
Lagrangian of Christ-Lee type with $L_{\rm{cl.}}(n)$ given by 
(2.37), ${\cal V}(n)$ by (2.47) and $L_{\rm{ext}}(n)$ 
by (3.10) at $\eta_n=\bar\eta_n=0$. 
One may introduce the ghost variables for the path integral (3.12), but 
they will be different from the ones in (3.9), 
since the argument of the jacobian ${\cal J}_n$ has been shifted from $y_n$  
to $\bar y_n$. On the other hand, the BRST identity can be 
constructed from the Slavnov-Taylor identity and we shall adapt this 
strategy. We consider a field dependent gauge transformation 
$$\delta r_n=-i\chi_n\sigma_2r_n,\eqno(3.14)$$
$$\delta z_n={1\over g}\chi_n,\eqno(3.15)$$
$$\delta\xi_n={1\over g}\dot\chi_n,\eqno(3.16)$$
where $$\chi_n={\varepsilon_n\over {1\over g}+\lambda y_n}\eqno(3.17)$$
with $\varepsilon_n$ an infinitesimal ordinary number.
Keeping in mind that the velocities $\dot x_n$, $\dot y_n$ and 
$\dot z_n$, and the coordinates $\bar x_n$, $\bar y_n$ and $\bar z_n$
follow strictly the discrete time definition and the variable 
transformation (3.14)-(3.16) is {\it{nonlinear}}, the variation of the 
Lagrangian $L_{\rm{cl.}}(n)$ contributes to the path integral  
in the limit $\epsilon\to 0$. We find that
$$\delta L_{\rm{cl.}}(n)={1\over 4}\epsilon^2g^2\xi_n\dot\chi_n
\widetilde{(Dr)_n}\dot r_n=-{1\over4}{\epsilon^2\lambda g^3
\xi_n\dot y_n\tilde{(Dr)_n}\dot r_n\over (1+\lambda g\bar y_n)^2}
\varepsilon_n,\eqno(3.18)$$ where the terms containing $\dot\varepsilon_n$ 
have been dropped since $\varepsilon_n$ is assumed smooth with respect 
to $n$ i.e. $\dot\varepsilon_n=O(1)$. Furthermore, the combination 
$dx_ndy_ndz_nd\xi_ndb_n\bar{\cal J}_n$ ceases to be invariant like 
the combination $dx_ndy_ndz_nd\xi_ndb_n{\cal J}_n$. We have, instead,
$$\delta\Big(dx_ndy_ndz_nd\xi_n\Big)=dx_ndy_ndz_nd\xi_n\Delta_n
$$ with $$\Delta_n=\biggl\{{1\over 2}\Big[{x_n\over(1+\lambda gy_n)^2}
-{x_{n+1}\over(1+\lambda gy_{n+1})^2}\Big]$$ $$+{1\over 4}
{\epsilon\lambda^2g^3\dot x_n\dot y_n\over (1+\lambda g\bar y_n)^3}
-{1\over 2}{\epsilon\lambda^3g^4\bar x_n\dot y_n^2\over(1+\lambda g
\bar y_n)^4}\biggr\}\varepsilon_n.\eqno(3.19)$$ With these additional 
terms, the identity (3.11) is replaced by $$\sum_n\varepsilon_n\Big[
\Big[-ig\tilde J_n\sigma_2<{1\over (1+\lambda gy_n)}r_n>+
\Big(\zeta_n-\dot u_n\Big)<{1\over 1+\lambda gy_n}>+<b_n>\Big]$$ 
$$+<\delta L_{\rm{cl.}}(n)-i\Delta_n-\delta{\cal V}(n)>\Big]
=0,\eqno(3.20)$$ where the term $<b_n>$ comes from the variation 
of the gauge fixing term of (3.13). Upon utilizing (2.45) and (2.46) for 
the Gauss average in the last section, we obtain that 
$$\sum_n\varepsilon_n<\delta L_{\rm{cl.}}(n)-i\Delta_n>=
\sum_n\varepsilon_n<\Big[-{1\over 8}\lambda^3g^4{\bar x_n\over
(1+\lambda g\bar y_n)^4}$$ $$+{3\over 8}{\lambda g^4(2+3\lambda^2)\bar x_n
-\lambda^2g^5(2+\lambda^2)\bar x_n\bar y_n\over (1+\lambda g\bar y_n)^5}
-{1\over 2}\lambda^3g^6(1+\lambda^2){\bar x_n^3\over 1+\lambda g\bar y_n)^6}
+O(\epsilon^{1\over 2})\Big]>$$ $$=\sum_n\varepsilon_n[<\delta{\cal V}
(n)>+O(\epsilon^{1\over 2})],\eqno(3.21)$$ and the Slavnov-Taylor 
identity follows in the limit $\epsilon\to 0$
$$-ig\tilde J_n\sigma_2<{1\over (1+\lambda gy_n)}r_n>+
(\zeta_n-\dot u_n)<{1\over 1+\lambda gy_n}>+<b_n>=0.\eqno(3.22)$$ 
This can also be obtained from (3.11) after integrating over $c$ 
and $\bar c$ at $\eta=\bar\eta=0$.

To construct the BRST identity, we introduce the ghost variables 
by rewriting (3.12) as 
$$e^{iW(J,\zeta,u,\eta,\bar\eta)}=\lim_{T\to\infty}
<\vert e^{-iHT}\vert>={\rm{const.}}\int\prod_{n=0}^N
dx_ndy_ndz_nd\xi_ndb_ndc_n^\prime d\bar c_n^\prime\times$$ 
$$\times <\vert x_N,y_N,z_N>e^{i\epsilon\sum_{n=0}^N[{\cal L}^\prime(n)
+L_{\rm{ext.}}^\prime(n)]}<x_0,y_0,z_0\vert>,\eqno(3.23)$$ where
$${\cal L}^\prime(n)=L_{\rm{cl.}}(n)+b_n(z_n-\lambda x_n)
-\bar c_n^\prime(1+\lambda g\bar y_n)
c_n^\prime-{\cal V}(n)\eqno(3.24)$$ and 
$$L_{\rm{ext.}}(n)=\tilde J_n r_n+\zeta_nz_n+u_n\xi_n+\bar
\eta_nc_n^\prime+\bar c_n^\prime\eta_n.\eqno(3.25)$$
Note that the primed ghosts are different from the original ones.
Denoting the average with respect to the path 
integral (3.23) by $<...>_\eta$, we have, for a function of the 
integration variables, $F$,  
$$<F>_\eta={<Fe^{i\epsilon\sum_n(\bar\eta_nc_n^\prime+\bar c_n^\prime
\eta_n)}>\over<e^{i\epsilon\sum_n(\bar\eta_nc_n^\prime+\bar c_n^\prime
\eta_n)}>}.\eqno(3.26)$$ Then it follows that
$$<c_n^\prime r_n>_\eta=<{1\over 1+\lambda g\bar y_n}r_n>\eta_n,
,\eqno(3.27)$$ $$<c_n^\prime>_\eta=<{1\over 1+\lambda g\bar y_n}>
\eta_n\eqno(3.28)$$ and $$<b_n>_\eta=<b_n>.\eqno(3.29)$$ 
In the limit $\epsilon\to 0$, the difference of $\bar y_n$ in (3.27) 
from $y_n$ may be neglected. The Slavnov-Taylor identity (3.22) implies the 
following BRST identity
$$-ig\tilde J_n\sigma_2<c_n^\prime r_n>_\eta+(\zeta_n
-\dot u_n)<c_n^\prime>_\eta+<b_n>_\eta\eta_n=0,\eqno(3.30)$$ which 
is equivalent to (3.11). As will be shown in  
Appendix B, the invariance of the 
Lagrangian under the field dependent transformation (3.14)-(3.17) 
is related to an symmetry of the corresponding Hamiltonian in the 
$\lambda$-gauge after factoring out the Gauss law constraint. There 
we shall present a derivation of the Slavnov-Taylor identity (3.22) 
from canonical formulations.

\section{IV. The Path Integral of an Nonabelian Gauge Field in 
Coulomb Gauge}

\subsection{IV.1 Quantization in the time axial gauge}

The Lagrangian density of a nonabelian gauge theory is 
$$L=-\int d^3\vec r\Big[{1\over 4}V_{\mu\nu}^lV_{\mu\nu}^l+\psi^\dagger
\gamma_4(\gamma_\mu D_\mu+m)\psi\Big],\eqno(4.1)$$ where 
$$V_{\mu\nu}^l={\partial V_\nu^l\over\partial x_\mu}
-{\partial V_\mu^l\over\partial x_\nu}+gf^{lmn}V_\mu^m V_\nu^n\eqno(4.2)$$
with $V_\mu^l$ the gauge potential and $f^{lmn}$ the structure constant 
of the Lie algebra of the gauge group. The fermion field $\psi$ carries 
both color and flavor indices and the mass matrix $m$ is diagonal with 
respect to the color indices. The gauge covariant derivative is 
$D_\mu={\partial\over\partial x_\mu}-igT^lV_\mu^l$ with $T^l$
the generator of the gauge group in the representation to which $\psi$
belongs. The normalizations of $f^{lmn}$ and $T^l$ are given by 
${\rm{tr}}T^lT^{l^\prime}={1\over 2}\delta^{ll^\prime}$ and 
$f^{lmn}f^{l^\prime mn}=C_2\delta^{ll^\prime}$ with $C_2$ the second 
Casmir of the gauge group.
The Lagrangian (4.1) is invariant under the following 
gauge transformation
$$V_\mu\to V_\mu^\prime=uV_\mu u^\dagger+{i\over g}u{\partial u^\dagger
\over\partial x_\mu}\eqno(4.3)$$ and 
$$\psi\to\psi^\prime=u\psi\eqno(4.4)$$ with $V_\mu=V_\mu^lT^l$ and 
$u$ the transformation matrix in the representation of $\psi$. 

The quantization of the gauge field is specified in the time axial 
gauge where $V_0=0$. The Lagrangian (4.1) becomes 
$$L=\int d^3\vec r\Big[{1\over 2}\dot V_j^l\dot V_j^l-{1\over 2}B_j^lB_j^l
+i\Psi^\dagger\dot\Psi-\Psi^\dagger\gamma_4(\gamma_j D_j+m)\Psi\Big],
\eqno(4.5)$$ and the corresponding Hamiltonian reads
$$H=\int d^3\vec r\Big[{1\over 2}\Pi_j^l\Pi_j^l+{1\over 2}
B_j^lB_j^l+\Psi^\dagger\gamma_4(\gamma_jD_j+m)\Psi\Big],\eqno(4.6)$$ 
where the canonical momentum 
$$\Pi_j^l(\vec r)=\dot V_j^l(\vec r)=-i{\delta\over\delta V_j^l(\vec r)}
\eqno(4.7)$$ and $B_j^l(\vec r)={1\over 2}\epsilon_{ijk}V_{jk}(\vec r)$ is 
the color magnetic field. The Hamiltonian (4.6) commutes with the 
generator of time-independent gauge transformations, ${\cal G}^l$
with $${\cal G}^l={1\over g}(\delta^{lm}\nabla_j-gf^{lmn}V_j^n)\Pi_j^m
+\Psi^\dagger T^l\Psi.\eqno(4.8)$$ A physical state in the Hilbert 
space is subject to the Gauss law constraint, i.e.
$${\cal G}^l\vert>=0.\eqno(4.9)$$ The path integral in the time axial 
gauge can be readily written down
$$<V\vert e^{-iHt}\vert>={\rm{const.}}\int\prod_n[dVd\Psi d\bar\Psi]_n
e^{i\epsilon\sum_n L(n)}<V_j(0,\vec r)\vert>,\eqno(4.10)$$ where
$$[dVd\Psi d\bar\Psi]_n\equiv\prod_{\vec r,j,l}
dV_j^l(n,\vec r)d\Psi(n,\vec r)d\bar\Psi(n,\vec r),\eqno(4.11)$$ 
$$L(n)=\int d^3\vec r\Big[{1\over 2}\dot V_j^l(n)\dot V_j^l(n)
-{1\over 2}B_j^l(n)B_j^l(n)$$ $$+i\bar\Psi(n)\gamma_4\dot\Psi(n)-\bar\Psi(n)
(\gamma_jD_j(n)+m)\Psi(n)\Big]\eqno(4.12)$$ with 
$$V_j^l(n)={V_j^l(n+1)-V_j^l(n)\over\epsilon}=O(\epsilon^{-{1\over 2}}),
\eqno(4.13)$$ and the initial 
wave functional satisfies the Gauss law (4.9). The dependence of the 
field amplitudes on $\vec r$ has been suppressed in (4.12).

\subsection{IV.2 Transformation to Coulomb Gauge}

Inserting the following identity into the path integral (4.10),
$$1={\rm{const.}}\int\prod_{n,\vec r}du(n,\vec r){\cal J}(n)
\delta(\nabla_jA_j^l(n,\vec r)),\eqno(4.14)$$ where 
$$A_j(n,\vec r)=u^\dagger(n,\vec r)V_j(n,\vec r)u(n,\vec r)
+{i\over g}u^\dagger(n,\vec r)\nabla_ju(n,\vec r),\eqno(4.15)$$
$u(n,\vec r)$ is a representation matrix of the gauge group and 
$${\cal J}(n)={\rm{det}}(-\nabla_j{\cal D}_j(n))\eqno(4.16)$$
with ${\cal D}_j^{lm}=\delta^{lm}\nabla_j-gf^{lmn}A_j^n$. Introducing 
$$u^\dagger(n,\vec r)u(n+1,\vec r)=e^{i\epsilon gA_0(n,\vec r)}
,\eqno(4.17)$$ $$\psi(n,\vec r)=u^\dagger(n,\vec r)\Psi(n,\vec r)
\eqno(4.18)$$ and $$\bar\psi(n,\vec r)=\bar\Psi(n,\vec r)
u(n+1,\vec r),\eqno(4.19)$$ and transforming the integration variables from 
$V_j^l(n,\vec r)$ and $u(n,\vec r)$ into $A_j^l(n,\vec r)$ and 
$A_0^l(n,\vec r)$, we obtain that
$$<V\vert e^{-iHt}\vert>={\rm{const.}}\int\prod_n
[dAd\psi d\bar\psi^\prime]_n\delta(\nabla_jA_j^l(n,\vec r))
e^{i\epsilon\sum_n L^\prime(n)}<A_j(0,\vec r)\vert>,\eqno(4.20)$$ where
$$L^\prime(n)=L(n)-{i\over\epsilon}\ln{\cal J}(n)-{i\over\epsilon}
\delta^3(0)\int d^3\vec r\ln h(n,\vec r)\eqno(4.21)$$ with $L(n)$ 
given by (4.12) and $h(n,\vec r)$ the Haar measure of the 
integration of the group element $e^{i\epsilon gA_0(n,\vec r)}$ with respect 
to $A_0(n,\vec r)$
$$h(n,\vec r)=1-{\epsilon^2g^2\over 24}A_0^l(n,\vec r)A_0^l(n,\vec r)
+O(\epsilon^3g^3),\eqno(4.22)$$ which does not have an analog in the 
soluble model. In terms of the new variables, we have
$$L(n)=\int d^3\vec r\Big[{\rm{tr}}[{\cal E}_j(n){\cal E}_j
(n)-{\cal B}_j(n){\cal B}_j(n)]$$ $${i\over\epsilon}
\bar\psi(n)[\psi(n+1)-e^{-i\epsilon gA_0(n)}
\psi(n)]-\bar\psi(n)e^{-i\epsilon gA_0(n)}
[\gamma_jD_j(n)+m]\psi(n)\eqno(4.23)$$ with 
$${\cal E}_j(n)=-{1\over\epsilon}\Big[e^{i\epsilon gA_0(n)}A_j(n+1)
e^{-i\epsilon gA_0(n)}-A_j(n)+{i\over g}e^{i\epsilon gA_0(n)}\nabla_j
e^{-i\epsilon gA_0(n)}\Big]\eqno(4.24)$$ and 
$${\cal B}_j(n)={1\over 2}\epsilon_{jki}\Big[\nabla_kA_i(n)
-\nabla_iA_k(n)-ig[A_k(n), A_i(n)]\Big].\eqno(4.25)$$
The Lagrangian (4.23) with (4.24) and (4.25) defines a gauge theory 
in a spatial continuum and on a temporal lattice. The action $\epsilon\sum_n
L(n)$ coincides with the naive continuum limit of the spatial links 
of Wilson's lattice action. But it comes naturally from the 
definition of a path integral and the procedure of gauge fixing. 
It follows from (4.13), (4.15) and (4.17) that [3] 
$$(\dot A_j^l(n,\vec r), A_0^l(n,\vec r))=O(\epsilon^{-{1\over 2}})
\eqno(4.26).$$ The path integral (4.20) can be rewritten as 
$$<V\vert e^{-iHt}\vert>={\rm{const.}}{\cal J}^{-{1\over 2}}(N)
\int\prod_n[dAd\psi d\bar\psi^\prime]_n$$ $$e^{i\epsilon\sum_n
L_{\rm{eff.}}(n)}{\cal J}^{{1\over 2}}(0)<A(0)\vert>\eqno(4.27)$$ with 
$$L_{\rm{eff.}}(n)=L(n)-{i\over\epsilon}[\ln\bar{\cal J}(n)+\ln h(n)]
+{i\over\epsilon}\Big[\ln\bar{\cal J}(n)-{1\over 2}(\ln{\cal J}(n+1)
+\ln{\cal J}(n))\Big],\eqno(4.28)$$ where 
$$\bar{\cal J}(n)={\rm{det}}(-\nabla_j\bar{\cal D}_j(n))\eqno(4.29)$$
with $\bar{\cal D}_j^{ab}=\delta^{ab}\nabla_j-gf^{abc}\bar A_j^c(n)$
and $\bar A_j^l(n)={1\over 2}[A_j^l(n+1)+A_j^l(n)]$. 
The small $\epsilon$ expansion of $L_{\rm{eff.}}(n)$ reads
$$L_{\rm{eff.}}(n)=L_{\rm{cl.}}(n)-{i\over\epsilon}\ln\bar{\cal J}(n) 
+\Delta L(n),\eqno(4.30)$$ where
$$L_{\rm{cl.}}(n)=\int d^3\vec r\Big[{\rm{tr}}[\bar{\cal E}_j(n)\bar{\cal E}_j
(n)-{\cal B}_j(n){\cal B}_j(n)]$$
$$+i\bar\psi(n)[\gamma_4(\dot\psi(n)+igA_0(n)\psi(n))-(\gamma_jD_j(n)+m)
\psi(n)]\Big]\eqno(4.31)$$ with 
$$\bar{\cal E}_j(n)=-\dot A_j(n)-\nabla_jA_0(n)-ig[A_0(n), \bar A_j(n)]
,\eqno(4.32)$$ and $$\Delta L(n)=\int d^3\vec r\Big[{1\over 8}g^2
\epsilon^2f^{lml^\prime}f^{akl^\prime}\bar{\cal E}_j^l(n)A_0^m(n)
A_0^k(n)\Big[\dot A_j^a(n)+{1\over 3}\bar{\cal D}_j^{ab}(n)A_0^b(n)\Big]
$$ $$-{i\over 8}\epsilon g^2(\vec r,l|[\nabla_j\bar{\cal D}_j(n)]^{-1}
t^m\dot A_{j^\prime}^m(n)\nabla_{j^\prime}[\nabla_i\bar{\cal D}_i(n)]^{-1}
t^{m^\prime}\dot A_{i^\prime}^{m^\prime}(n)\nabla_{i^\prime}|\vec r,l)$$ $$
+{i\over24}\epsilon\delta^3(0)C_2g^2A_0^l(n)A_0^l(n)+{i\over 2}\epsilon
g^2\bar\psi(n)\gamma_4T^lT^m\psi(n)A_0^l(n)A_0^m(n)\Big]
+O(\epsilon^{1\over 2})\eqno(4.33)$$ 
with $t^l$ the generator in the adjoint representation, $(t^l)^{ab}=if^{alb}$. 
The first term of the integrand of $\Delta L(n)$ comes from the 
$\epsilon$ expansion of the color electric field (4.24), the second term
from the shift of the Jacobian ${\cal J}(n)$, the last term of (4.28), 
the third term comes from the Haar measure $h$ and the last term from the 
$\epsilon$ expansion of the fermionic part of (4.23). We may notice the close 
resemblance of the first two terms of (4.33) with (2.38). 

\subsection{IV.3 Converting $\Delta L$ into an equivalent potential}

Following the recipe of Section III, the potential energy which is 
equivalent to $\Delta L(n)$ in the limit $\epsilon\to 0$ is 
$${\cal V}=-<\Delta L(n)>_{\rm{Gauss}}$$ $$\equiv-{\int\prod_{\vec r,j,l}
d\dot A_j^l(n,\vec r)dA_0^l(n,\vec r)\delta(\nabla_jA_j^l(n,\vec r))
e^{i\epsilon\int d^3\vec r{\rm{tr}}\bar{\cal E}_j(n,\vec r)
\bar{\cal E}_j(n,\vec r)}\Delta L(n)\over\int\prod_{\vec r,j,l}
d\dot A_j^l(n,\vec r)dA_0^l(n,\vec r)\delta(\nabla_jA_j^l(n,\vec r))
e^{i\epsilon\int d^3\vec r{\rm{tr}}\bar{\cal E}_j(n,\vec r)
\bar{\cal E}_j(n,\vec r)}}\eqno(4.34)$$ while regarding $\bar A_j(n,\vec r)$
constant. The Gauss average of a product of $\dot A_j$ and $A_0$ can be 
decomposed by Wick's theorem. We have
$$<{\cal E}_i^a(\vec r){\cal E}_j(\vec r^\prime)>_{\rm{Gauss}}={i\over\epsilon}
\delta_{ij}(\vec r,a|\vec r^\prime,b)={i\over\epsilon}\delta_{ij}
\delta^{ab}\delta^3(\vec r-\vec r^\prime),\eqno(4.35)$$
$$<A_0^a(\vec r)A_0^b(\vec r^\prime)>_{\rm{Gauss}}=-{i\over\epsilon}
(\vec r,a|G\nabla^2G|\vec r^\prime,b),\eqno(4.36)$$
$$<A_0^a(\vec r)\dot A_j^b(\vec r^\prime)>_{\rm{Gauss}}=-{i\over\epsilon}
\Big[(\vec r,a|G\nabla_j|\vec r^\prime,b)+(\vec r,a|G\nabla^2G
{\cal D}_j|\vec r^\prime,b)\Big],\eqno(4.37)$$
$$<\dot A_i^a(\vec r)\dot A_j^b(\vec r^\prime)>_{\rm{Gauss}}=
{i\over\epsilon}\Big[\delta_{ij}\delta^{ab}\delta^3(\vec r-\vec r^\prime)
+(\vec r,a|\nabla_iG{\cal D}_j|\vec r^\prime,b)$$
$$+(\vec r,a|{\cal D}_iG\nabla_j|\vec r^\prime,b)
+(\vec r,a|{\cal D}_iG\nabla^2G{\cal D}_j|\vec r^\prime,b)\Big],\eqno(4.38)$$
where we have suppressed the $n$-dependence and 
$G=(-\nabla_j{\cal D}_j)^{-1}$ with ${\cal D}$ from here on to the end of the 
section defined at 
$\bar A_j(n,\vec r)$. Substituting (4.35)-(4.38) into (4.33), we obtain 
$${\cal V}=-{1\over 24}C_2g^2\delta^3(0)\int d^3\vec r(\vec r,l|
G\nabla^2G|\vec r,l)$$ $$+{1\over 8}g^2f^{kam}f^{nal}\int d^3\vec r
(\vec r,l|G\nabla_j|\vec r,k)(\vec r,m|G\nabla_j|\vec r,n)$$
$$-{1\over 4}g^2f^{kam}f^{nbl}\int d^3\vec r\int d^3\vec r^\prime
(\vec r,l|G\nabla_i|\vec r^\prime,k)(\vec r,n|\nabla_jG|\vec r^\prime,m)
(\vec r,b|{\cal D}_jG\nabla_i|\vec r^\prime,a)$$
$$+{1\over 8}g^2f^{kam}f^{nbl}\int d^3\vec r\int d^3\vec r^\prime
(\vec r,l|G\nabla_i|\vec r^\prime,k)(\vec r^\prime,m|G\nabla_j|\vec r,n)
(\vec r^\prime,a|{\cal D}_iG\nabla^2G{\cal D}_j|\vec r,b)$$ $$
+{3\over 8}C_2g^2\delta^3(0)\int d^3\vec r(\vec r,m|G\nabla^2G|\vec r,m)
$$ $$+{1\over 8}g^2f^{lka}f^{mna}\int d^3\vec r(\vec r,k|G\nabla_j|\vec r,l)
(\vec r,m|G\nabla_j|\vec r,n)$$
$$-{1\over 8}g^2f^{nka}f^{lma}\int d^3\vec r(\vec r,k|G\nabla_j|\vec r,l)
(\vec r,m|G\nabla_j|\vec r,n)$$
$$+{1\over 12}g^2f^{lmk}f^{ank}\int d^3\vec r\Big[
(\vec r,m|G\nabla_j|\vec r,l)(\vec r,a|{\cal D}_jG\nabla^2G|\vec r,n)$$
$$+(\vec r,a|{\cal D}_jG\nabla_j|\vec r,l)(\vec r,m|G\nabla^2G|\vec r,n)
+(\vec r,n|G\nabla_j|\vec r,l)(\vec r,a|{\cal D}_jG\nabla^2G|\vec r,m)
$$ $$-{1\over 2}g^2\int d^3\vec r(\vec r,l|G\nabla^2G|\vec r,m)\bar\psi
(\vec r)\gamma_4T^lT^m\psi(\vec r).\eqno(4.39)$$
The first term is the Wick contraction of the Haar measure term of (4.33), 
the second to the fourth terms are from the jacobian of the gauge fixing, 
i.e., the second term of (4.33), the fifth to the eighth terms are 
from the color electric field
energy, i.e., the first term of (4.33) and the last term is from the 
fermion part. This lengthy expression can be simplified with the aid 
of the following two Jacobian identities, i.e. 
$$f^{abc}\int d^3\vec r[(\vec r,a|{\cal D}_i|X)(\vec r,b|Y)(\vec r,c|Z)
+(\vec r,a|X)(\vec r,b|{\cal D}_j|Y)(\vec r,c|Z)$$
$$+(\vec r,a|X)(\vec r,b|Y)(\vec r,c|{\cal D}_j|Z)]=0,\eqno(4.40)$$
[3] and $$f^{abl}f^{ckl}+f^{bcl}f^{akl}+f^{cal}f^{bkl}=0.\eqno(4.41)$$
First of all, the seventh term of (4.39) is already of the form of 
Christ-Lee's ${\cal V}_1$. The covariant derivative ${\cal D}_j$ of the 
third term may be moved into the middle factor of the integrand according 
to (4.40), and the result will cancel with the second and the sixth terms 
through (4.41). Upon repeat applications of (4.40) and (4.41), the first, 
forth, fifth and eighth terms will combine into Christ-Lee's ${\cal V}_2$. 
We have finally
$${\cal V}={\cal V}_1+{\cal V}_2+{\cal V}_3,\eqno(4.42)$$
where $${\cal V}_1={1\over 8}g^2\int d^3\vec r(\vec r,l^\prime|
G\nabla_j|\vec r,l)(\vec r,m|G\nabla_jt^{l^\prime}t^l|\vec r,m),
\eqno(4.43)$$ $${\cal V}_2=-{1\over 8}g^2\int d^3\vec r
\int d^3\vec r^\prime(\vec r^\prime,l^\prime|(\delta_{i^\prime i}
+{\cal D}_{i^\prime}G\nabla_i)|\vec r,n)
(\vec r,l|(\delta_{ii^\prime}+{\cal D}_iG\nabla_{i^\prime})|\vec r^\prime,
n^\prime)$$ $$\times (\vec r,n|t^lG\nabla^2Gt^{l^\prime}|\vec r^\prime,
n^\prime)\eqno(4.44)$$ and $${\cal V}_3=-{1\over 2}g^2\int d^3\vec r
\bar\psi(\vec r)\gamma_4T^lT^m\psi(\vec r)
(\vec r,l|G\nabla^2G|\vec r,m).\eqno(4.45)$$
The terms ${\cal V}_1$ and ${\cal V}_2$ are the Christ-Lee operator  
ordering terms for a pure gauge theory. The term ${\cal V}_3$ is new and its
expansion in $g$ reads
$${\cal V}_3=-{1\over 2}g^2\int d^3\vec r\Big[\delta^{lm}
(\vec r|\nabla^{-2}|\vec r)+3g^2f^{ll^\prime k}f^{knm}
(\vec r|\nabla^{-2}A_i^{l^\prime}\nabla_i\nabla^{-2}A_j^n\nabla_j\nabla^{-2}
|\vec r)$$ $$+O(g^3A^3)\Big]\bar\psi(\vec r)\gamma_4T^lT^m
\psi(\vec r),\eqno(4.46)$$ where the term linear in $A_j^l$ vanishes because 
of the transversality. Operatorwise, this term stems from the normal 
ordering of the four fermion coupling in the color Coulomb potential, 
which is necessary for the passage from the canonical formulation to the 
path integral. The details will be explained in Appendix C.
The effective Lagrangian in the path 
integral (4.27) is replaced by the following Lagrangian of Christ-Lee 
type in the limit $\epsilon\to 0$
$${\cal L}(n)=L_{\rm{cl.}}(n)-{i\over\epsilon}\ln\bar{\cal J}(n) 
-{\cal V}_1(n)-{\cal V}_2(n)-{\cal V}_3(n),\eqno(4.47)$$

The formulation of this section for the Coulomb gauge can be easily 
generalized to an arbitrary noncovariant gauge introduced in [3]
$$\int d^3\vec r^\prime(\vec r,l|\Gamma_j|\vec r^\prime,l^\prime)
A_j^{l^\prime}(\vec r^\prime)=0.\eqno(4.48)$$

Since $\epsilon$ is the only dimensional parameter in the formal manipulation
of this section, it would be expected that $A_0(n,\vec r)=O(\epsilon^{-1})$ 
on dimensional grounds, different from the estimate of $\xi$ for the 
soluble model and the estimate (4.26). On the other hand, the field theory 
in $D=4$ suffers from the ultraviolet divergences which have to be 
regularized in order for the path integral to make sense. The validity of 
the estimate $A_0(n,\vec r)=O(\epsilon^{-{1\over 2}})$ as well as the 
Christ-Lee path integral depends on an implicit assumption that there is an 
fixed ultraviolet length, which makes the summation over all physical degrees 
of freedom finite, in the process $\epsilon\to 0$. If $\epsilon$ is 
identified with the ultraviolet length as in the discrete time regularization
scheme of the next section, the $\epsilon$-expansion can nolonger be truncated. 

\section{V. The BRST Identity and the Discrete Time Regularization}

Neglecting fermion couplings, the Lagrangian of a nonabelian gauge field 
with discrete times reads
$$L(n)=\int d^3\vec r{\rm{tr}}[{\cal E}_j(n){\cal E}_j
(n)-{\cal B}_j(n){\cal B}_j(n)]\eqno(5.1)$$ with ${\cal E}_j$ and 
${\cal B}_j$ given by (4.24) and (4.25). The corresponding path 
integral is $$<V\vert e^{-iHt}\vert>={\rm{const.}}\int\prod_n 
[dAdbdcd\bar c]_ne^{i\epsilon\sum_n L_{\rm{BRST}}(n)}
<A_j(0)\vert>,\eqno(5.2)$$ where $$[dAdbdcd\bar c]_n
=\prod_{\vec r,\mu,l}dA_\mu^l(n,\vec r)db^l(n,\vec r)
dc^l(n,\vec r)d\bar c^l(n,\vec r)h(n,\vec r)\eqno(5.3)$$ and
$$L_{\rm{BRST}}(n)=L(n)+\int d^3\vec rb^l(n,\vec r)\nabla_jA_j^l(n,\vec r)
-\int d^3\vec r\bar c^l(n,\vec r)[\nabla_j{\cal D}_j
(n,\vec r)]^{ll^\prime}c^{l^\prime}(n,\vec r).\eqno(5.4)$$ 
The Lagrangian (5.4) and the integration measure of (5.3) are invariant 
under the following transformation:
$$\delta A_j^l(n,\vec r)=s_n{\cal D}_j^{ll^\prime}(n,\vec r)c^{l^\prime}
(n,\vec r),\eqno(5.5)$$ $$\delta e^{i\epsilon gA_0(n,\vec r)}
=igs_nc^l(n,\vec r)T^le^{i\epsilon gA_0(n,\vec r)}-igs_{n+1}c^l(n+1,\vec r) 
e^{i\epsilon gA_0(n,\vec r)}T^l,\eqno(5.6)$$
$$\delta c^l(n,\vec r)=-{1\over 2}s_ngf^{lab}c^a(n,\vec r)c^b(n,\vec r),
\eqno(5.7)$$ $$\delta\bar c^l(n,\vec r)=s_nb^l(n,\vec r)\eqno(5.8)$$ and 
$$\delta b^l(n,\vec r)=0,\eqno(5.9)$$ where $s_n$ is a Grassmann 
number. For a $n$-independent $s_n$, a nilpotent charge operator can be 
extracted and the transformation (5.5)-(5.9) is 
therefore of BRST type. Introducing the generating 
functional of the connected Green's functions via a source term, i.e.
$$e^{iW(J,\eta,\bar\eta)}=\lim_{T\to\infty}
<\vert e^{-iHT}\vert>$$ $$={\rm{const.}}\int\prod_n[dAdbdcd\bar c]_n
<\vert A(N)>e^{i\epsilon\sum_n[L_{\rm{BRST}}(n)+L_{\rm{ext.}}(n)]}
<A(0)\vert>\eqno(5.10)$$ with 
$$L_{\rm{ext.}}(n)=2\int d^3\vec r{\rm{tr}}[J_\mu(n,\vec r)A_\mu(n,\vec r)
+\bar\eta(n,\vec r)c(n,\vec r)+\bar c(n,\vec r)\eta(n,\vec r)].
\eqno(5.11)$$ The invariance of (5.3) and (5.4) under (5.5)-(5.9) 
implies the following BRST identity [11]
$$\int d^3\vec r{\rm{tr}}[\vec J(n,\vec r)\cdot<\vec {\cal D}c(n,\vec r)>
-<{\cal D}_0J_0(n,\vec r)>+ig\bar\eta(n,\vec r)<c^2(n,\vec r)>$$
$$+<b(n,\vec r)>\eta(n,\vec r)]=0,\eqno(5.12)$$ where 
$$\vec {\cal D}c(n,\vec r)=\vec\nabla c(n,\vec r)
-ig[\vec A(n,\vec r),c(n,\vec r)]
\eqno(5.13)$$ and $${\cal D}_0J_0(n,\vec r)=\dot J_0(n,\vec r)
+ig[A_0(n,\vec),J_0(n,\vec r)].\eqno(5.14)$$ The transformation law 
of $A_0(n,\vec r)$, deduced from (5.6), 
$$\delta A_0(n,\vec r)=-{\cal D}_0\theta(n,\vec r)+{1\over 12}g^2\epsilon^2
[A_0(n,\vec r),[A_0(n,\vec r),\dot\theta(n,\vec r)]]+...\eqno(5.15)$$
has been utilized, only the first term of which contributes to the limit
$\epsilon\to 0$. The identity (5.12) can be cast into 
various useful forms [11].

Similar to the case of the soluble model, the BRST identity can also be
constructed from the Slavnov-Taylor identity of the Christ-Lee type of 
path integral (4.27) with $L_{\rm{eff.}}(n)$ replaced by ${\cal L}(n)$ 
of (4.47) in the limit $\epsilon\to 0$. 

Unlike the soluble model, the field theory case suffers from an ultraviolet 
divergence which needs to be regularized and subtracted. Owing to its 
manifest BRST invariance, the discrete time Lagrangian (4.23) with 
(4.24) and (4.25) serves also as a gauge invariant regularization scheme
with $\epsilon$ a ultraviolet cutoff. There are several additional technical 
advantages with this regularization. 1) The energy integration with 
a continuum time is regularized by the summation over the Bloch momentum 
on the temporal lattice. This is particularly important for resolving the 
ambiguities associated with the energy divergence. 
2) With fixed Bloch momenta on 
the temporal lattice, the integration over spatial momenta is less 
divergent. There is only a finite number of divergent skeletons and these 
can be handled by the dimensional regularization; 3) For fixed lattice 
momenta, the integrand of each Feynman diagram is a rational function 
of the spatial momenta and can be simplified with the aid of Feynman 
parametrization; 4) Manifest unitarity is maintained throughout the 
calculation. In what follows, we shall test this regularization by an 
evaluation of the one loop Coulomb propagator in the absence of the quark 
fields. 

The expansion of the Lagrangian (5.3) according to the power of $g^2$ reads
$$L_{\rm{BRST}}(n)=L_{\rm{cl.}}(n)+\int d^3\vec rb^l(n,\vec r)
\nabla_jA_j^l(n,\vec r)$$ $$-\int d^3\vec r\bar c^l(n,\vec r)
[\nabla_j{\cal D}_j(n,\vec r)]^{ll^\prime}c^{l^\prime}(n,\vec r)
+R_n,\eqno(5.16)$$ where
$$R_n=\int d^3\vec r\Big[-{1\over 8}g^2
\epsilon^2f^{lml^\prime}f^{akl^\prime}\Big(\dot A_j^l(n)A_0^m(n)
A_0^k(n)\dot A_j^a(n)$$ $$+{1\over 3}\nabla_jA_0^l(n)A_0^m(n)A_0^k(n)
\nabla_jA_0^a(n)\Big)+{i\over24}\epsilon\delta^3(0)C_2g^2A_0^l(n)A_0^l(n)
\Big]\eqno(5.17)$$ where at the order $g^2$, only terms of an even number 
of $A_0$ factor are kept. 
The first term of (5.17) comes from the expansion of $e^{i\epsilon g
A_0}$ in the color electric field and the second from the Haar measure. 
Both of them have been included in $\Delta L(n)$ of (4.33). For the 
reason we shall explain later, the perturbative expansion ought to 
be performed in Euclidean space, which amounts to  
the substitutions $\epsilon\to -i\epsilon$, 
$A_0\to -iA_4$ and $\dot A_j\to i{\partial A_j\over \partial x_4}$. 
The dressed Coulomb propagator reads
$$d_0^\prime(k_0,\vec k)={1\over \vec k^2+\sigma(k_0,\vec k)},$$
where the one loop contribution to $\sigma(k_0,\vec k)$ is given by the 
amputated Feynman diagrams of Fig. 1. plus the contribution of (5.17), i.e.  
$$\sigma(k_0,\vec k)=-\Big(\hbox{ Fig. 1a }+\hbox{ Fig. 1b }+\hbox{ Fig. 1c }
\Big)+\hbox{ contribution from $R_n$}\eqno(5.18)$$ with the relevant 
Feynman rules given in Fig. 2. A wavy line stands for a 
transverse gluon propagator and contributes a factor

\topinsert
\hbox to\hsize{\hss
	\epsfxsize=4.0truein\epsffile{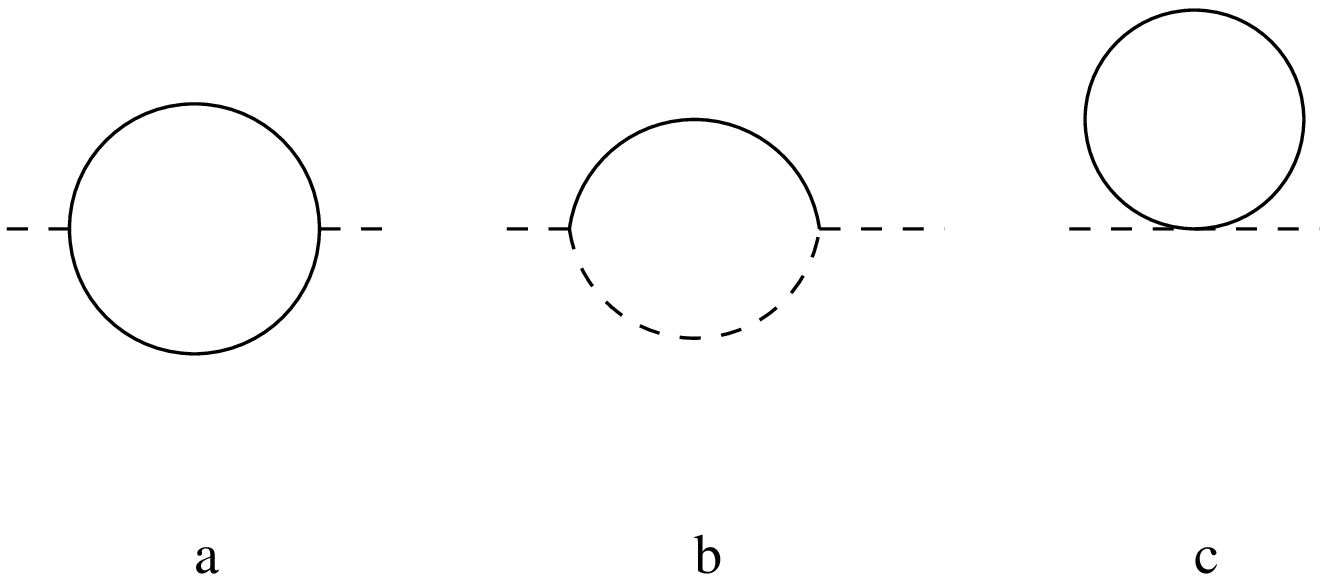}
	\hss}
\begincaption{Figure 1}
The one loop diagrams of the inverse Coulomb propagator.
\endcaption
\endinsert
\topinsert
\hbox to\hsize{\hss
	\epsfxsize=4.0truein\epsffile{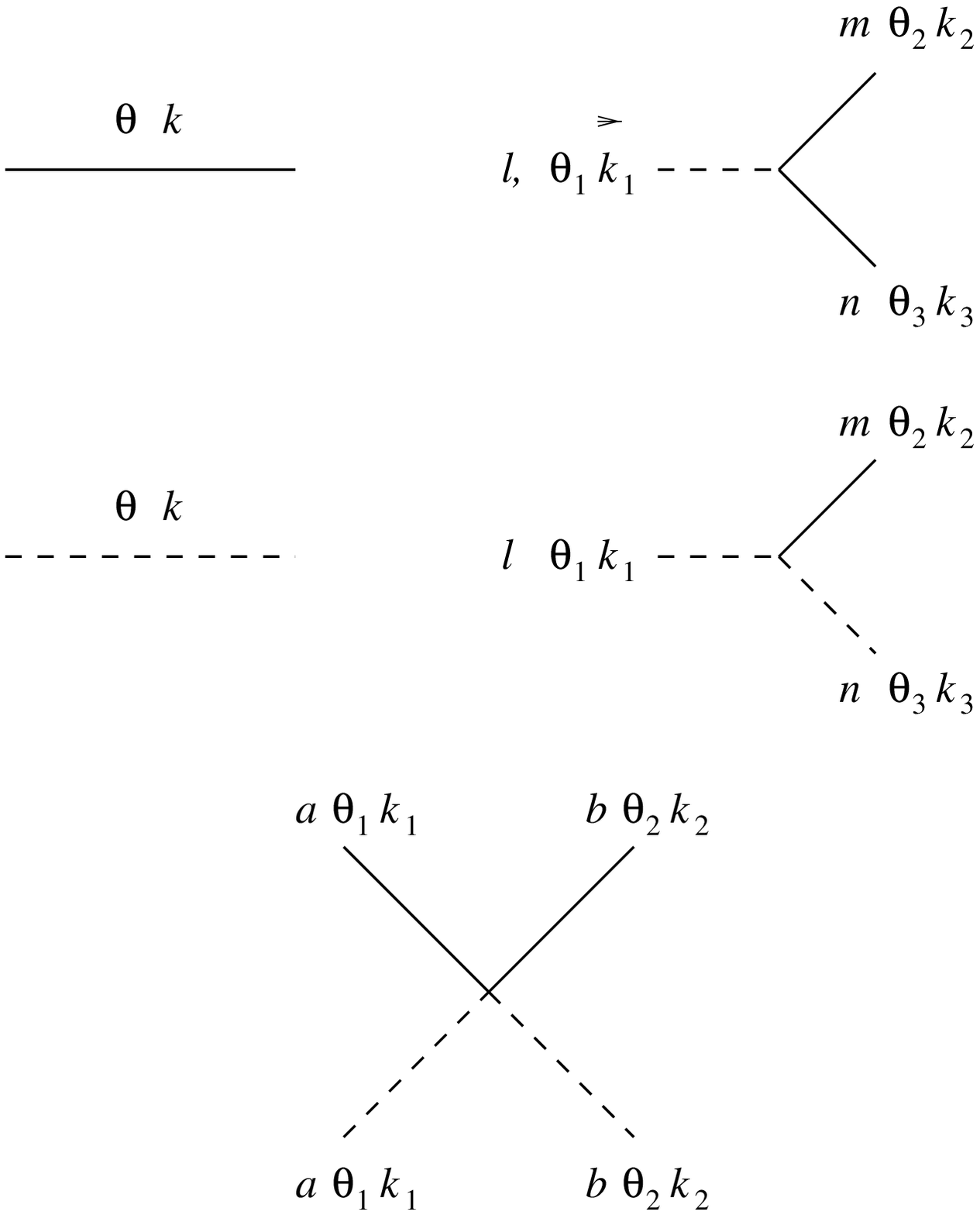}
	\hss}
\begincaption{Figure 2}
The relevant ingredients of the diagrams
\endcaption
\endinsert

$$\delta^{ll^\prime}d_{ij}(\theta|\vec k)={\delta^{ll^\prime}
\over k_0^2+\vec k^2}\Big(\delta_{ij}-
{k_ik_j\over \vec k^2}\Big)\eqno(5.19)$$ with $k_0={2\over\epsilon}
\sin{\theta\over 2}$ and $\theta\in (-\pi,\pi)$ a Bloch momentum; a dashed 
line stands for a bare Coulomb propagator and contributes a factor
$$\delta^{ll^\prime}d_0(\vec k)={\delta^{ll^\prime}\over \vec k^2}.
\eqno(5.20)$$ A three point vertex of one Coulomb line and two transverse 
gluons with incoming momenta $(\theta_1,\vec k_1)$, $(\theta_2,\vec k_2)$ and 
$(\theta_3,\vec k_3)$ is associated with the factor
$$-i{2\over\epsilon}gf^{lmn}\delta_{ij}\sin{\theta_3-\theta_2\over 2};
\eqno(5.21)$$ a three point vertex of two Coulomb lines and one 
transverse gluon with incoming momenta $(\theta_1,\vec k_1)$, 
$(\theta_2,\vec k_2)$ and $(\theta_3,\vec k_3)$ is associated with the 
factor $$-igf^{lmn}(k_{2j}-k_{1j})\cos{\theta_3\over 2}.\eqno(5.22)$$
A four point vertex of two transverse gluons and two Coulomb lines 
is associated with the factor
$$-g^2(f^{la^\prime a}f^{lb^\prime b}+f^{la^\prime b}f^{lb^\prime a})
\cos{\theta_1\over 2}\cos{\theta_2\over 2}\delta_{ij}.\eqno(5.23)$$ 
With these rules, we have
$$\hbox{ Fig.1a }={1\over 2}C_2\delta^{ll^\prime}g^2{4\over\epsilon^2}
\int_{-\pi}^\pi{d\theta\over 2\pi\epsilon}
\sin^2{\theta+\theta^\prime\over 2}I,\eqno(5.24)$$ where
$$I=\int{d^3\vec p\over (2\pi)^3}d_{ij}(\theta|\vec p)
d_{ij}(\theta^\prime|\vec p^\prime)$$
$$=3!\int{d^3\vec p\over (2\pi)^3}\int_0^1dx\int_0^1dy\int_0^1dz
z(1-z){\vec p^2(\vec p+\vec k)^2+[(\vec p\cdot(\vec p+\vec k)]^2
\over[(\vec p+\vec kz)^2+p_0^2x(1-z)+p_0^{\prime2}yz+\vec k^2z(1-z)]^4}
\eqno(5.25)$$ with $p_0={2\over\epsilon}\sin{\theta\over 2}$,
$p_0^\prime={2\over\epsilon}\sin{\theta^\prime\over 2}$ and 
$\phi=\theta^\prime-\theta$, $\vec k=\vec p^\prime-\vec p$ the 
external energy and momentum. Similarly,
$$\hbox{ Fig. 1b }=C_2\delta^{ll^\prime}g^2\int_{-\pi}^\pi
{d\theta\over 2\pi\epsilon}\cos^2{\theta\over 2}II,\eqno(5.26)$$ where
$$II=\int{d^3\vec p\over (2\pi)^3}(p+2k)_i(p+2k)_j
d_{ij}(\theta|\vec p)d_0(\vec p^\prime)$$
$$=8\int{d^3\vec p\over (2\pi)^3}\int_0^1dx\int_0^1dz(1-z)
{\vec p^2\vec k^2-(\vec p\cdot\vec k)^2\over [(\vec p+\vec kz)^2
+p_0^2x(1-z)+\vec k^2z(1-z)]^3}\eqno(5.27)$$ and 
$$\hbox{ Fig. 1c }=-C_2\delta^{ll^\prime}g^2\int_{-\pi}^\pi
{d\theta\over 2\pi\epsilon}\cos^2{\theta\over 2}
\int{d^3\vec p\over (2\pi)^3}d_{jj}(\theta|\vec p)$$
$$=-2C_2\delta^{ll^\prime}g^2\int_{-\pi}^\pi
{d\theta\over 2\pi\epsilon}\cos^2{\theta\over 2}
\int{d^3\vec p\over (2\pi)^3}{1\over p_0^2+\vec p^2}\eqno(5.28)$$
and $$\hbox{Contribution of $R_n$}=-C_2\delta^{ll^\prime}{g^2\over 12}
\int_{-\pi}^\pi{d\theta\over 2\pi\epsilon}\int{d^3\vec p\over(2\pi)^3}
\int_0^1dx{(24\sin^2{\theta\over 2}+\vec k^2\epsilon^2)\vec p^2+p_0^2
\vec k^2\epsilon^2\over (\vec p^2+p_0^2x)^2}.\eqno(5.29)$$ 
We shall not expose the details of the evaluation of (5.25)-(5.29), but 
only remark on few key points which lead to the final answer. 
First of all, the $\vec p$-integrations in (5.27)-(5.29)
are all linearly divergent, which upon the replacement
$$\int {d^3\vec p\over (2\pi)^3}\to \int {d^D\vec p\over (2\pi)^D}
,\eqno(5.30)$$ give rise to Gamma functions with arguments of
the form ${D\over 2}+\hbox{integer}$, and therefore yield finite limits as 
$D\to 3$. After the $\vec p$-integration, the integrand for $\theta$-
integration is of the dimension of a momentum. Because of the $\epsilon$ 
of the denominator of the Bloch momentum $p_0$, the leading divergence 
as $\epsilon\to 0$ is of the order of $\epsilon^{-2}$, which reflects the 
usual quadratic divergence. At $\vec k=0$, we obtain that 
$$\hbox{(5.24)}={2\over 3\pi^2\epsilon^2}C_2g^2\delta^{ll^\prime},
\eqno(5.31)$$ $$\hbox{(5.26)}=0,\eqno(5.32)$$
$$\hbox{(5.28)}={2\over 3\pi^2\epsilon^2}C_2g^2\delta^{ll^\prime},
\eqno(5.33)$$ and $$\hbox{(5.29)}={4\over 3\pi^2\epsilon^2}C_2g^2
\delta^{ll^\prime}.\eqno(5.34)$$ If follows from (5.18) that 
$$\sigma(k_0,\vec k)=0,\eqno(5.35)$$ which renders the net divergence
logarithmic. After some manipulations, we obtain that
$$\sigma(0,\vec k)=-{11\over 24\pi^2}C_2g^2\vec k^2\Big(\ln{1\over 
k\epsilon}-{74\over 33}-{91\over 22}\ln2\Big)\eqno(5.36)$$ and the 
one loop renormalized Coulomb propagator reads
$$d_0^\prime(\vec k)={Z\over \vec k^2}\eqno(5.37)$$ with 
$$Z=1+{11\over 24\pi^2}C_2g^2\Big(\ln{1\over 
k\epsilon}-{74\over 33}-{91\over 22}\ln2\Big),\eqno(5.38)$$
the divergent part of which coincides with the charge renormalization
[1] [2].

We end this section with two technical remarks:

1). Euclidean time is adapted for the above one loop calculation.
This turns out to be necessary for the logarithmically divergent 
diagrams with the integration order we followed. Consider a simple integral 
with a Minkowski momentum $p=(p_0,\vec p)$
$$I=\int {d^4p\over (2\pi)^4}{i\over (p^2+m^2)^2}\eqno(5.39)$$ with 
$p^2=\vec p^2-p_0^2$. If the Wick rotation is performed before the 
spatial integration, the infinite arc, $p_0=Re^{i\phi}$ with $R\to\infty$, 
$0<\phi<{\pi\over 2}$ and $\pi<\phi<{3\pi\over 2}$ will not contribute. 
But if the spatial momentum is integrated first as we did, the Wick rotation 
then will pick up a term from the infinite arc. As a result, the 
renormalization constant will be complex unless we start with the Euclidean 
definition of the diagram.

2). It may looks puzzling that the very terms of $R_n$ which help to cancel 
the quadratic divergence of the one loop diagrams of the Coulomb propagator 
are actually the same terms which contribute to the Christ-Lee anomalous 
vertices which are expected at two loop level. This paradox is tied to the 
identification $\epsilon$ with the ultraviolet cutoff. 
If an independent  ultraviolet cutoff is introduced 
for the integration over $\vec p$ and the limit $\epsilon\to 0$ is taken 
before sending the cutoff to infinity, the contribution of $R_n$ to the 
one loop Coulomb propagator, (5.28), will vanish as can be seen easily.

\section{VI. Concluding Remarks}

In this work, we have carefully traced all the subtleties of gauge 
fixing and variable transformation in a path integral of a gauge 
model, without resorting to the operator 
formalism. For a soluble quantum mechanical model in $\lambda$-
gauge and for a nonabelian gauge field in Coulomb gauge, the well 
known operator ordering terms are reproduced exactly. In the presence 
of fermionic degrees of freedom, an additional operator ordering term
is discovered. Because of the intrinsic nonlinearity of a BRST 
transformation, the operator ordering terms are found essential in restoring 
the simple form of the identity associated with this transformation. In the 
field theory case, a manifest BRST invariant and unitary regularization 
scheme is proposed and it does give rise to the correct $\beta$-function 
at one loop order.

Though this work does not attempt to prove the renormalizability 
of a nonabelian gauge theory in Coulomb gauge, I do not see any problems 
in applying the discrete time regularization scheme to higher orders.
The only draw back is that the $\epsilon$-expansion of the temporal 
lattice Lagrangian can no longer be truncated since the ultraviolet 
cutoff is identified with $\epsilon$.   

Alternatively, one may try to renormalize the theory with a Christ-Lee 
type of path integral. Then one has to face the energy divergence and the 
ambiguities associated with it. The coupling with the ultraviolet 
divergence makes it difficult to organize the cancellation in higher orders. 
Several scenarios have been proposed but none of them [12] goes smoothly 
beyond two loops. On the other hand, the energy divergence is an 
artifact of the path integral, since it is not there with canonical 
perturbation methods. In principle, one should be able to organize the 
energy integral before integrating spatial momenta and to reproduce
the canonical perturbation series. But then no advantages of Feynman 
diagrams have been taken and the path integral seems unnecessary. 
What we need for the renormalization with a Christ-Lee type of path
integral is an unambiguous scheme which regularize the spatial loop integral. 
The only feasible BRST invariant scheme is a spatial lattice. 

At this point, it is instructive to draw some connections of the path 
integral in the continuum with Wilson's lattice formulation [13]. In the 
absence of quarks, the partition function of Wilson's formulation on  
a four dimensional rectangular lattice reads
$$Z=\int\prod_{<ij>}dU_{ij}e^{-{1\over g^2}S_W[U]},\eqno(6.1)$$
where $U_{ij}$ is a gauge group matrix on a nearest neighbor link. 
The simplest choice of the action is
$$S_W[U]={1\over g^2}\Big[{a_s\over a_t}\sum_{P_t}{\rm{tr}}\Big(1-{1\over d}
{\rm{Re}}U_{P_t}\Big)+{a_t\over a_s}\sum_{P_s}{\rm{tr}}\Big(1-{1\over d}
{\rm{Re}}U_{P_s}\Big)\Big],\eqno(6.2)$$ where $U_P=U_{ij}U_{jk}
U_{kl}U_{li}$ for a plaquette $P(ijkl)$ with the subscript $s$ labeling 
the space-like one and $t$ time-like one, $a_s$, $a_t$ denote 
the spatial and temporal lattice spacings and $d$ the dimension of 
$U$'s. The lattice Coulomb gauge 
condition can be imposed as in Ref. 14. The discrete time regularization 
scheme presented in Section 5 corresponds to the limit 
$a_t\to 0$ after $a_s\to 0$ and any  regularization corresponding 
to Christ-Lee path integral follow from the limit $a_s\to 0$ after 
$a_t\to 0$. 

\section{Acknowledgments}

The author would like to thank Professor D. Zwanziger for illuminating 
discussions. he is also obliged to Professor N. N. Khuri and Dr. James Liu 
for their reading of the manuscript. This work is supported in part by U. S. 
Department of Energy under Grant DE-FG02-91ER40651, Task B and by National 
Science Council of ROC under Grant NSC-CTS-981003. 

The author would like to dedicate this work to late Ms. Irene Tramm. Her
selfless help to his career is highly appreciated. 

\section{Appendix A}

To estimate the typical contributions of $\dot x$, $\dot y$, $\dot z$ 
and $\xi$ to the path integral in the limit $\epsilon\to 0$,  
we may neglect the interaction term and consider the path integral with 
the free Lagrangian only, 
$$L_0(n)={1\over 2}[\dot x_n^2+\dot y_n^2+(\dot z_n-\xi_n)^2
-(\omega^2-i0^+)(x_n^2+y_n^2)-{1\over a}
(z_n-\lambda x_n-\kappa\dot\xi_n)^2],
\eqno(A.1)$$ where the total time interval $T=N\epsilon\to\infty$ with 
$N$ the number of time slices between the time interval $T$, and the 
infinitesimal imaginary part of $\omega$ provides a converging factor of the 
integral. The last term of (A.1) is the gauge fixing term (2.49) with the 
gauge parameter $a$. Defining the path integral average of an arbitrary 
function of $x_n$, $y_n$, $z_n$ and $\xi_n$ by 
$$<F>={\int\prod_ndx_ndy_ndz_nd\xi_ne^{i\epsilon\sum_nL_0(n)}F\over
\int\prod_ndx_ndy_ndz_nd\xi_ne^{i\epsilon\sum_nL_0(n)}}.\eqno(A.2)$$
we obtain the following expressions for various propagators:
$$<x_nx_m>=<y_ny_m>={1\over\epsilon}\int_{-\pi}^\pi {d\theta\over 2\pi}
{ie^{-i(n-m)\theta}\over p^*p-\omega^2+i0^+},\eqno(A.3)$$ 
$$<z_nz_m>={1\over\epsilon}\int_{-\pi}^\pi{d\theta\over 2\pi} 
i{\lambda^2-(a-\kappa^2p^*p)(p^*p-\omega^2)\over (1+\kappa p^2)
(1+\kappa p^{*2})(p^*p-\omega^2+i0^+)}e^{-i(n-m)\theta}\eqno(A.4)$$
$$<\xi_n\xi_m>={1\over\epsilon}\int_{-\pi}^\pi{d\theta\over 2\pi}
i{\lambda^2p^*p+(1-ap^*p)(p^*p-\omega^2)\over (1+\kappa p^2)
(1+\kappa p^{*2})(p^*p-\omega^2+i0^+)}e^{-i(n-m)\theta}\eqno(A.5)$$
$$<x_n\xi_m>=-{1\over\epsilon}\int_{-\pi}^\pi{d\theta\over 2\pi}
{\lambda p^*e^{-i(n-m)\theta}\over (1+\kappa p^{*2})(p^*p-\omega^2+i0^+)},
\eqno(A.6)$$ $$<x_nz_m>={1\over\epsilon}\int_{-\pi}^\pi{d\theta\over 2\pi}
{\lambda e^{-i(n-m)\theta}\over (1+\kappa p^{*2})(p^*p-\omega^2+i0^+)}
\eqno(A.7)$$ and $$<z_n\xi_m>=-{1\over\epsilon}\int_{-\pi}^\pi
{d\theta\over 2\pi}i{\lambda^2p^*
-(\kappa p+ap^*)(p^*p-\omega^2)\over (1+\kappa p^2)
(1+\kappa p^{*2})(p^*p-\omega^2+i0^+)}e^{-i(n-m)\theta},\eqno(A.8)$$
where $p=i{e^{-i\theta}-1\over\epsilon}$. According to the definition
of $\dot x_n$ and $\dot y_n$, we have
$$<\dot x_n\dot x_m>=<\dot y_n\dot y_m>={1\over\epsilon}
\int_{-\pi}^\pi {d\theta\over 2\pi}{ip^*p
e^{-i(n-m)\theta}\over p^*p-\omega^2+i0^+}.\eqno(A.9)$$ 

The squares of the typical magnitude of $x_n$, $y_n$, $z_n$, $\dot x_n$, 
$\dot y_n$, $\dot z_n$ and $\xi_n$ inside the path integral in the limit 
$\epsilon\to 0$ are of the same order as the expectation value of 
their squares, i.e. the propagators (A.3)-(A.5) at $n=m$. It follows 
from (A.3) and (A.9) that $$<x_n^2>=<y_n^2>={1\over 2\omega}
+O(\epsilon^2),\eqno(A.10)
$$ but $$<\dot x_n^2>=<\dot y_n^2>={i\over\epsilon}+{\omega\over 2}
+O(\epsilon^2)\eqno(A.11)$$ for arbitrary $\lambda$, $\kappa$ and 
$a$. On the other hand, the $\epsilon\to 0$ limit of $<z_n^2>$,
$<\dot z_n^2>$ and $<\xi_n^2>$ are very delicate and we consider the 
following situations. 

1) None of $\kappa$ or $a$ vanishes, It 
follows from (A.4) and (A.5) that
$$<z_n^2>={\lambda^2\over 2\omega(1+\kappa\omega^2)^2}
+{i\over\sqrt{\kappa}}\Big[\kappa-a-{\lambda^2\kappa\over
1+\kappa\omega^2}-{2\lambda^2\kappa\over(1+\kappa\omega^2)^2}\Big]
+O(\epsilon^2),\eqno(A.12)$$
$$<\xi_n^2>={\lambda^2\omega\over2(1+\kappa\omega^2)^2}
+{i\over4\kappa\sqrt{\kappa}}\Big[\kappa-a-{\lambda^2\kappa\over 
1+\kappa\omega^2}+{2\lambda^2\kappa\over(1+\kappa\omega^2)^2}
\Big]+O(\epsilon^2)\eqno(A.13)$$ and 
$$<\dot z_n^2>={i\over\epsilon}+\hbox{finite terms}.\eqno(A.14)$$
The small $\epsilon$ expansion of the lattice Lagrangian (2.27) is trivial
with such a gauge fixing. So is the case when $x_n$ and $z_n$ in the last term
of (A.1) are replaced by $\bar x_n$ and $\bar z_n$. 

2) $\kappa=0$ and $a\to 0$ before $\epsilon\to 0$. 
This is the $\lambda$-gauge in the text. It is easy to show, using 
(A.3), (A.4) and (A.5) that
$$<z_n^2>=\lambda^2<x_n^2>={\lambda^2\over 2\omega},\eqno(A.15)$$ 
$$<\dot z_n^2>=\lambda^2<\dot x_n^2>={i\lambda^2\over\epsilon}+
\hbox{finite terms}\eqno(A.16)$$ and
$$<\xi_n^2>={i(1+\lambda^2)\over\epsilon}+\hbox{finite terms}.\eqno(A.17)$$

3) $\kappa=0$ but $a\neq 0$. This corresponds to the 
``smeared $\lambda$-gauge''. We obtain from (A.4) and (A.5) that
$$<z_n^2>=-i{a\over\epsilon}+\hbox{finite terms},\eqno(A.18)$$
$$<\dot z_n^2>=-2i{a\over\epsilon^2}+i{\lambda^2\over\epsilon}
+\hbox{finite terms}\eqno(A.19)$$ and 
$$<\xi_n^2>=-2i{a\over\epsilon^3}+i{1+\lambda^2\over\epsilon}
+\hbox{finite terms}.\eqno(A.20)$$ The equations (A.12)-(A.20) 
give the announced estimates in Section II.

\section{Appendix B}

The Hamiltonian of the soluble model in the $\lambda$-gauge is given by [5]
$$H={1\over 2{\cal J}}\left(\matrix{p_x&p_y}\right){\cal J}\left(
\matrix{{\cal M}_{xx}^{-1}&{\cal M}_{xy}^{-1}\cr{\cal M}_{yx}^{-1}
&{\cal M}_{yy}^{-1}}\right)\left(\matrix{p_x\cr  p_y}\right)
+U(x^2+y^2)\eqno(B.1)$$ after solving the Gauss law constraint, where 
$${\cal M}_{xx}^{-1}={\cal J}^{-2}(y^2+{1\over g^2}),\eqno(B.2)$$
$${\cal M}_{xy}^{-1}={\cal M}_{yx}^{-1}={\cal J}^{-2}x({\lambda\over g}
-y),\eqno(B.3)$$ 
$${\cal M}_{yy}^{-1}={\cal J}^{-2}\Big[\Big(\lambda y+{1\over g}\Big)^2
+x^2(\lambda^2+1)\Big]\eqno(B.4)$$ and 
$${\cal J}={1\over g}+\lambda y.\eqno(B.5)$$
It was pointed out that the Hamiltonian (B.1) commutes with the operator
$$K={\cal J}^{-1}(xp_y-yp_x),\eqno(B.6)$$ i.e. $[H,K]=0$ [5]. With ${\cal U}
=e^{i\varepsilon K}$, we have 
$$x_\varepsilon\equiv{\cal U}x{\cal U}^{-1}=x-\varepsilon{\cal J}^{-1}y
\eqno(B.7)$$ and 
$$y_\varepsilon\equiv{\cal U}y{\cal U}^{-1}=y+\varepsilon{\cal J}^{-1}x
\eqno(B.8)$$ for infinitesimal $\varepsilon$. Adding the source term
$$h(t)=\kappa(t)K+J_x(t)x(t)+J_y(t)y(t)\eqno(B.9)$$ to the 
Hamiltonian (B.1), the Schroedinger equation of the state is given 
by $$i{\partial\over\partial t}|t>=h(t)|t>,\eqno(B.10)$$ where the 
operators follow the time development generated by $H$, e.g.
$$x(t)=e^{iHt}x(0)e^{-iHt}\eqno(B.11)$$ and 
$$y(t)=e^{iHt}y(0)e^{-iHt}.\eqno(B.12)$$ The $c$-number sources $\kappa(t)$,
$J_x(t)$ and $J_y(t)$ are adiabatically switched on in the remote past and 
are switched off in the remote future. It can be shown that
$$\Big({\cal J}^{1\over 2}K{\cal J}^{-{1\over 2}}\Big)_W=
{\cal J}^{1\over 2}K{\cal J}^{-{1\over 2}}.\eqno(B.13)$$
with the subscript $W$ standing for the Weyl ordering.
The general solution of (B.10) reads $$|t>=U(t,t_0)|t_0>\eqno(B.13)$$ with 
$$U(t,t_0)=T\exp\Big(-i\int_{t_0}^tdt^\prime h(t^\prime)\Big).\eqno(B.14)$$
Define the generating functional of the connected Green's functions, 
${\cal W}(\kappa,J)$ by   
$$e^{i{\cal W}(\kappa,J)}=<|U(\infty,-\infty)|>\eqno(B.15)$$ with $|>$ the
ground state of the Hamiltonian (B.1), the previously defined one, 
$W(J,\zeta,u,\eta,\bar\eta)$ in (3.12) at $\zeta=u=\eta=\bar\eta=0$
corresponds to ${\cal W}(0,J)$. Effecting an infinitesimal 
transformation (B.7) and (B.8) with $\varepsilon(t)\to 0$ at 
$t\to\pm\infty$, we have 
$$i{\partial\over\partial t}|t>_\varepsilon=h_\varepsilon(t)|t>_\varepsilon,
\eqno(B.16)$$ where
$$|t>_\varepsilon={\cal U}(t)|t>\eqno(B.17)$$ and 
$$h_\varepsilon(t)=(\kappa-{\partial\varepsilon\over\partial t})K
+J_x(t)x_\varepsilon(t)+J_y(t)y_\varepsilon(t).\eqno(B.18)$$  Consequently, 
$$|t>_\varepsilon=U_\varepsilon(t,t_0)|t_0>_\varepsilon\eqno(B.19)$$ with 
$$U_\varepsilon(t,t_0)=T\exp\Big(-i\int_{t_0}^tdt^\prime 
h_\varepsilon(t^\prime)
\Big).\eqno(B.20)$$ The invariance of the Hamiltonian $H$ and its 
ground state under the transformation implies that
$$<|U_\varepsilon(\infty,-\infty)-U(\infty,-\infty)|>=0,\eqno(B.21)$$ 
which, to the linear power of $\varepsilon$ gives
$$\int_\infty^\infty dt\biggl\{{\partial\varepsilon\over\partial t}
<K>_t+\varepsilon(t)\Big[J_x(t)<{gy\over 1+\lambda gy}>_t-J_y(t)
<{gx\over 1+\lambda gy}>_t\Big]\biggr\}=0,\eqno(B.22)$$ where the 
canonical average $<...>_t$ is defined as $$<O>_t
={<|T[U(\infty,-\infty)O(t)]|>\over<|U(\infty,-\infty)|>}.\eqno(B.23)$$
Converting $<|U(\infty,-\infty)|>$ into the path integral and denoting 
the path integral average by $<...>$ without the subscript $t$, we 
have $$<{gy\over 1+\lambda gy}>_t=<{gy(t)\over 1+\lambda gy(t)}>,
\eqno(B.24)$$ $$<{gx\over 1+\lambda gy}>_t=<{gx(t)\over 1+\lambda gy(t)}>
\eqno(B.25)$$ and $$<K>_t|_{\kappa=0}=-g
<{\dot x(t)y(t)-x(t)\dot y(t)+\lambda g[x^2(t)+y^2(t)]\dot x(t)
\over 1+g^2[x^2(t)+y^2(t)]}>\vert_{\kappa=0}.\eqno(B.26)$$ 
The last equality requires some explanation. In the canonical form, 
we may write $$<K>_t|_{\kappa=0}=-{\delta\over\delta\kappa(t)}
{\cal W}(\kappa,J)|_{\kappa=0}.\eqno(B.27)$$ On the other hand, the 
operator $K$ contains the canonical momenta. Performing a Legendre 
transformation of the Hamiltonian $H+h$ [15], the term of the corresponding 
Lagrangian which is linear in $\kappa$ reads $$\kappa g{\dot xy-x\dot y+\lambda 
g(x^2+y^2)\dot x\over 1+g^2(x^2+y^2)}\eqno(B.28)$$ and the equality (B.26) 
follows from (B.27) with the path integral representation of 
${\cal W}(\kappa,J)$. Putting back the $\xi$ and $z$, we find 
$$<K>_t|_{\kappa=0}=<\xi(t)-\dot z(t)>\eqno(B.29)$$ and the identity (B.22) 
becomes $$\int_\infty^\infty dt\biggl\{{\partial\varepsilon\over\partial t}
<\xi(t)-\dot z(t)>+\varepsilon(t)\Big[J_x(t)<{gy(t)\over 1+\lambda gy(t)}>
-J_y(t)<{gx(t)\over 1+\lambda gy(t)}>\Big]\biggr\}=0,\eqno(B.30)$$
For an arbitrary function $\varepsilon(t)$, the integration sign may be 
removed after a partial integral and the Slavnov-Taylor identity 
(3.22) at $u=\zeta=0$ emerges. The relation 
$$<b(t)>={d\over dt}<\xi(t)-\dot z(t)>,\eqno(B.31)$$ 
which can be checked explicitly, is utilized in the final step.

\section{Appendix C}

To make the paper self-contained, we shall go through the path integral 
of fermionic degrees of freedom, following the coherent field 
treatment of the Ref. [7].
Consider a pair of fermion annihilation and creation operators, $a$,
and $a^\dagger$, with anticommutator
$$\{a, a^\dagger\}=1,\eqno(C.1)$$ The combination $a^\dagger aa^\dagger a$ 
is not zero. But if we replace $a$ and $a^\dagger$ by a pair of 
Grassmann numbers $z$ and $\bar z$, the combination $\bar zz\bar zz$ 
is always zero. Therefore there are ordering ambiguities when transforming 
the canonical formulation for fermionic operators to the path integral. 
The question is which order goes through to the path integral 
simply through the above replacements $a\to z$ and $a^\dagger\to\bar z$. 
We shall discuss the systematics in the following:

For a system of $M$ fermionic degrees of freedom, represented by the 
annihilation and creation operators $a_j$ and $a_j^\dagger$ with
$$\{a_i, a_j\}=0\eqno(C.2)$$ and $$\{a_i, a_j^\dagger\}=\delta_{ij}
,\eqno(C.3)$$ we introduce two set of independent Grassmann 
numbers, $z_1$, $z_2$,..., $z_M$ and $\bar z_1$, $\bar z_2$,..,
$\bar z_M$. We also specify that they anticommute with the $a$'s, 
$a^\dagger$'s and commute with the ket or bra of the ground state in 
the Hilbert space. Furthermore, the following integration rule is imposed
$$\int dz_j=\int d\bar z_j=0\eqno(C.4)$$ and 
$$\int z_idz_j=\int d\bar z_i\bar z_j=\delta_{ij}.\eqno(C.5)$$ 
Defining a coherent state by
$$|z_1,z_2,...,z_M>\equiv e^{\sum_ja_j^\dagger z_j}|0>\eqno(C.6)$$
and its conjugate by 
$$<\bar z_1,\bar z_2,...,\bar z_M|\equiv <0|e^{\sum_j\bar z_ja_j}\eqno(C.7)$$
with $|0>$ the ground state. It follows that
$$a_j|z_1,z_2,...,z_M>=z_j|z_1,z_2,...,z_M>,\eqno(C.8)$$ 
$$<\bar z_1,\bar z_2,...,\bar z_M|a_j^\dagger=<\bar z_1,\bar z_2,...
\bar z_M|\bar z_j\eqno(C.9)$$ and 
$$<\bar z_1,\bar z_2,...,\bar z_M|z_1,z_2,...,z_M>=e^{\sum_j\bar z_jz_j}.
\eqno(C.10)$$ Furthermore, we have the completeness relation
$$\int|z_1,z_2,...,z_M>\prod_jdz_jd\bar z_j e^{-\sum_j\bar z_jz_j}<\bar z_1,
\bar z_2,...,\bar z_M|=1.\eqno(C.11)$$ Let the Hamiltonian of the 
system be $$H(a^\dagger,a)=\sum_{ij}\omega_{ij}a_i^\dagger a_j+{1\over 2}
\sum_{ii^\prime,jj^\prime}v_{ii^\prime,j^\prime j}
a_i^\dagger a_{i^\prime}^\dagger a_{j^\prime}a_j+...$$ 
$$+{1\over M!}\sum_{i_1,...,i_M;j_M,...,j_1}v_{i_1...i_M,j_M...j_1}
a_{i_1}^\dagger...a_{i_M}^\dagger a_{j_M}...a_{j_1},\eqno(C.12)$$
where the normal ordering with respect to the state $|0>$ is the 
crucial point. It follows from (C.8) and (C.9) that
$$<\bar z_1,...,\bar z_M|H|z_1,...,z_M>=H(\bar z,z)$$ $$=
\sum_{ij}\omega_{ij}\bar z_iz_j+{1\over 2}
\sum_{ii^\prime,jj^\prime}v_{ii^\prime,j^\prime j}
\bar z_i\bar z_{i^\prime}z_{j^\prime}z_j+...$$ $$+{1\over M!}
\sum_{i_1,...,i_M;j_M,...,j_1}\bar z_{i_1}...\bar z_{i_M}z_{j_M}
...z_{j_1}\eqno(C.13)$$ and therefore 
$$<\bar z_1,...,\bar z_M|e^{-i\epsilon H}|z_1,...,z_M>
=e^{\sum_j\bar z_jz_j}[1-i\epsilon H(\bar z,z)+O(\epsilon^2)].
\eqno(C.14)$$ 
With the aid of the completeness relation (C.11), we end up with the 
following path integral representation of the fermionic system
$$<\bar z_1^\prime,...,\bar z_M^\prime|e^{-itH}|z_1,...,z_M>
=\int[dz]_N\prod_{n=1}^{N-1}[d\bar zdz]_n[d\bar z]_0e^{i\sum_n
L(n)}\eqno(C.15)$$ where $t=N\epsilon$ and $\epsilon\to 0$ at fixed 
$t$, and we have made the abbreviation 
$$[d\bar zdz]_n=\prod_jd\bar z_j(n)dz_j(n),\eqno(C.16)$$ 
$$[dz]_N=\prod_jdz_j(N)\eqno(C.17)$$ and $$[d\bar z]_0=\prod_jd\bar z_j(0)
\eqno(C.18)$$ The Lagrangian $L(n)$ reads $$L(n)=i\sum_j\bar z_j(n)\dot z_j(n)
-H(\bar z(n),z(n))\eqno(C.19)$$ with $\dot z_j(n)={1\over\epsilon}
[z_j(n+1)-z_j(n)]$.

Like bosonic operators, the ordering ambiguity here is also reflected 
in the difference between the Dyson-Wick contraction and the path 
integral contraction at equal time. Consider a free system whose 
Hamiltonian is given by (C.12) with $\omega_{ij}=\omega\delta_{ij}$ and all 
$v$'s vanishing. The Dyson-Wick contraction gives. 
$$\lim_{t\to 0^+}<0|T(a(t)a^\dagger(0))|0>=1\eqno(C.20)$$ while
$$\lim_{t\to 0^-}<0|T(a(t)a^\dagger(0))|0>=0.\eqno(C.21)$$ The path 
integral, on the other hand, gives rise to an unambiguous result 
at $t=0$ since
$$S_{ij}\equiv{\int [dz]_N\prod_{n=1}^{N-1}[d\bar zdz]_n
[d\bar z]_0z_i(m)\bar z_j(m)e^{i\epsilon\sum_nL(n)}\over
\int [dz]_N\prod_{n=1}^{N-1}[d\bar zdz]_n
[d\bar z]_0e^{i\epsilon\sum_nL(n)}}$$ $$
=\delta_{ij}{1\over\epsilon}\int_{-\pi}^\pi{d\theta\over 2\pi}{i\over 
i{e^{-i\theta}-1\over\epsilon}-\omega+i0^+}=0.\eqno(C.22)$$

To illustrate the caution which is needed in transforming the 
canonical formulation to path integral, we consider a soluble 
gauge model whose Lagrangian is given by
$$L={1\over 2}(\dot z-\xi)^2+i\psi^\dagger(\dot\psi-ig\xi)
-m\psi^\dagger\psi\eqno(C.23)$$ with $\psi$, $\psi^\dagger$ fermionic 
and $z$, $\xi$ bosonic. The gauge transformation reads
$$z\to z^\prime=z+{\alpha\over g},\eqno(C.24)$$ 
$$\xi\to\xi^\prime=\xi+{\dot\alpha\over g}\eqno(C.25)$$
and $$\psi\to\psi^\prime=e^{i\alpha}\psi\eqno(C.26)$$ with $\alpha$ an 
arbitrary function of time. In the time axial gauge where $\xi=0$,
The Hamiltonian corresponding to (C.23) is 
$$H=-{1\over 2}{\partial^2\over \partial Z^2}+m\Psi^\dagger\Psi
\eqno(C.27)$$ and the Gauss law constraint is
$$\Big(-i{\partial\over\partial Z}-g\Psi^\dagger\Psi\Big)|>=0
\eqno(C.28)$$ The constraint can be solved explicitly and the 
physical spectrum consists of two states with $\Psi^\dagger\Psi
=0,1$ and the corresponding eigenvalue of $H$ $=0$ and $m+{g^2\over 2}$.
Though trivial, we still follow the transformation of (C.27) and (C.28) to 
the gauge where $z=0$, with the dynamical variables $\theta$ determined 
by $Z+{\theta\over g}=0$ and $\psi=e^{i\theta}\Psi$. The Hamiltonian
(C.27) and the constraint (C.28) becomes
$$H=-{g^2\over 2}{\partial^2\over \partial\theta^2}+m\psi^\dagger\psi
\eqno(C.29)$$ and
$$\Big(-i{\partial\over\partial\theta}-\psi^\dagger\psi\Big)|>=0.\eqno(C.30)$$
Substituting the solution of (C.30) into (C.29), we obtain 
$$H_{\rm{eff.}}=m\psi^\dagger\psi+{g^2\over 2}(\psi^\dagger\psi)^2.
\eqno(C.31)$$ 
Following the above recipe, we convert (C.28) into a path integral
$$<\psi|e^{-itH_{\rm{eff.}}}|>=\int\prod_ndz_nd\xi_nd\psi_n
d\bar\psi_n\delta(z_n)e^{i\epsilon\sum_nL_{\rm{eff.}}(n)}<\psi_0|>
,\eqno(C.32)$$ where
$$L_{\rm{eff.}}(n)={1\over 2}(\dot z_n-\xi_n)^2+i\bar\psi_n(\dot\psi_n
-ig\xi_n\psi_n)-m\bar\psi_n\psi_n-{g^2\over 2}\bar\psi_n\psi_n
\eqno(C.33)$$ with the last term comes from the normal ordering of the 
four-fermion term of (C.31). The 
integration over $\xi_n$ in (C.33) will not generate quartic terms 
since the combination $(\bar\psi_n\psi_n)^2$ vanishes. Applying the 
Feynman rules given by the path integral (C.32) at $t\to
\infty$, we have verified explicitly that the shift of the self-energy 
because of the interaction vanishes to one loop order, 
in agreement with the result of canonical quantization.

Finally, we come to the nonabelian gauge field. The four fermion 
Coulomb interaction term of Christ-Lee Hamiltonian in Coulomb 
gauge reads
$$H_{\rm{Coul.}}={g^2\over 2}\int d^3\vec rd^3\vec r^\prime
\psi^\dagger(\vec r)T^l\psi(\vec r)(\vec r,l|G(-\nabla^2)G|
\vec r^\prime,l^\prime)\psi^\dagger(\vec r^\prime)T^{l^\prime}
\psi(\vec r^\prime)$$ $$={g^2\over 2}\int d^3\vec rd^3\vec r^\prime
:\psi^\dagger(\vec r)T^l\psi(\vec r)(\vec r,l|G(-\nabla^2)G|
\vec r^\prime,l^\prime)\psi^\dagger(\vec r^\prime)T^{l^\prime}
\psi(\vec r^\prime):$$ $$+{g^2\over 2}\int d^3\vec r 
(\vec r,l|G(-\nabla^2)G|\vec r^\prime,l^\prime)\psi^\dagger(\vec r)
T^lT^{l^\prime}\psi(\vec r).\eqno(C.34)$$ The last term becomes 
${\cal V}_3$ of (4.45). 

The additional term steming from the normal ordering of fermionic 
operators begins to show up at one loop level, unlike its
bosonic counterpart. In the case of an abelian 
gauge theory, the term (4.45) corresponds to the self Coulomb 
energy of a fermion and is not observable, but here, for the nonabelian 
case, it carries the coupling to the gluon fields and may not be 
ignored.

\bigskip
\references

\ref{1.}{J. Frenkel and J. C. Taylor, {\it Nucl. Phys.} {\bf B109}, 439
(1976).}             

\ref{2.}{T. D. Lee, {\it Particle Physics and Introduction to Field Theory},
Harwood Academic, Chur, 1981.}

\ref{3.}{N. H. Christ and T. D. Lee, {\it Phys. Rev.} {\bf D22}, 939(1980).}

\ref{4.}{H. Cheng and E. C. Tsai, {\it Phys. Rev. Lett.} {\bf 57}, 511
(1986).}

\ref{5.}{R. Friedberg, T. D. Lee, Y. Pang and H. C. Ren, {\it Ann. Phys.} 
{\bf Vol. 246}, 381(1996).}

\ref{6.}{J.-L. Gervais and A. Jevicki, {\it Nucl. Phys.}, {\bf B110}, 93(1976).
See also the footnote 9 of the Ref. 3}

\ref{7.}{B. Sakita, {\it Quantum Theory of Many-Variable Systems and Fields},
World Scientific, 1985.} 

\ref{8.}{R. P. Feynman and A. R. Hibbs, {\it Quantum Mechanics and Path 
Integrals}, McGraw-Hill, New York, 1965.}

\ref{9.}{K. Fujikawa, {\it Nucl. Phys.} {\bf B468}, 355 (1996).}

\ref{10.}{A. A. Slavnov, {\it Theor. Math. Phys.} {\bf 10}, 99 (1972); 
J. C. Taylor, {\it Nucl. Phys.} {\bf B33}, 436 (1971).}

\ref{11.}{D. Zwanziger, {\it Nucl. Phys}. {\bf B518}, 237(1998); 
L. Baulieu and D. Zwanziger, hep-th/9807024.}

\ref{12.}{H. Cheng and E. C. Tsai, {\it Chinese J. Phys.} {\bf 25}, 1(1987);
P. Doust, {\it Ann. Phys.} {\bf Vol. 177}, 169(1987); P. J. Doust and J. C. 
Taylor, {\it Phys. Lett.} {\bf 197B}, 232(1987).}

\ref{13.}{K. G. Wilson, {\it Phys. Rev.} {\bf D14}, 2455(1974).}

\ref{14.}{D. Zwanziger, {\it Nucl. Phys.} {\bf B485}, 185(1997).}

\ref{15.}{T. D. Lee and C. N. Yang, {\it Phys. Rev.} {\bf 128}, 885(1962).}

\vfill\eject
\end